\documentclass[epj]{svjour}
\usepackage{graphicx}
\usepackage{amsmath}
\usepackage{hyperref}
\usepackage{multirow}
\usepackage{subcaption}
\usepackage{color}
\begin{document}
\title{Impedance modelling and collective effects in the Future Circular e$^+$e$^-$ Collider with 4 IPs\footnote{This work was partially supported by the European Union’s Horizon 2020 research and innovation programme under grant No 951754 - FCCIS Project, by the National Natural Science Foundation of China, Grant No. 11775238, and by INFN National committee V through the ARYA project.}}
\author{M. Migliorati\inst{1}\thanks{\emph{Present address:} mauro.migliorati@uniroma1.it}, C. Antuono\inst{1,2}, E. Carideo\inst{1,2}, Y. Zhang\inst{3} \and M. Zobov\inst{4}
%
}                     
%
%
\institute{University of Rome ‘La Sapienza’ and INFN Sezione Roma1, 00185 Roma, Italy \and CERN, 1217 Meyrin, Geneva, Switzerland \and Institute of High Energy Physics, 100049 Beijing, China \and INFN - LNF, Frascati, 00044 Roma, Italy}
\date{Received: date / Revised version: date}

\abstract{
The FCC-ee impedance model is being constantly updated closely following the vacuum chamber design and parameters evolution. In particular, at present, a thicker NEG coating of 150 nm (instead of previous 100 nm) has been suggested by the vacuum experts, and a more realistic impedance model of the bellows has been investigated. Moreover, also the transverse impedance has been updated by considering the same sources as for the longitudinal case. Therefore, the FCC-ee impedance database is getting more complete and the impedance model is being refined.
In this paper we describe the presently available machine coupling impedance in both longitudinal and transverse planes, and study the impedance-driven single bunch instabilities (with and without beam-beam interaction) for the new FCC-ee parameter set with
4 interaction points (IPs). The results are compared with the previously obtained ones and a further possible mitigation of the beam-beam head-tail instability (X-Z instability) is proposed.
\PACS{
      {PACS-key}{discribing text of that key}   \and
      {PACS-key}{discribing text of that key}
     } 
} 
\maketitle
\section{Introduction}
\label{intro}

The Future Circular Collider (FCC) project is an ambitious program that comprises, in a single tunnel of about 100 km, both hadron~\cite{FCChh} (FCC-hh) and electron-positron~\cite{fcc} (FCC-ee) colliders. The challenging machines should be located in the CERN area and should represent the future of the particle physics in the post-LHC (Large Hadron Collider) era.

FCC-ee will operate with four different energies, 45.6, 80, 120 and 182.5 GeV, optimised to study with high precision the Z, W, Higgs and top particles, respectively. The project is now in the phase of developing the feasibility study in order to provide input for the next European Particle Physics Strategy Update, which will be held in 2026-2027.

The parameter list of all the four options of FCC-ee, updated with respect to the conceptual design report (CDR)~\cite{fcc} is shown in Table~\ref{tab:1}. In particular the high bunch population and small emittances make the operation at 45.6 Gev the most critical for collective effects and instabilities. Their mitigation is, therefore, one of the most important tasks to be solved in order to achieve the desired design performance. The beam impedance related instabilities~\cite{fccee1}, \cite{fccee2}, \cite{fccee3}, \cite{ipac21}, electron cloud effects~\cite{fccee5}, \cite{Yaman}, ion induced phenomena~\cite{Mether} and other harmful collective effects need to be thoroughly studied to accomplish this task.

\begin{table}
\centering
\caption{Parameter list used in simulations}
\label{tab:1}       
\begin{tabular}{|l|c|c|c|c|}
\hline
Layout & \multicolumn{4}{c|}{PA31-1.0} \\
 & Z & WW & ZH & t$\hat{\mathrm{t}}$ \\
\hline
Circumference (km) & \multicolumn{4}{c|}{91.174117 km} \\
\hline
Beam energy (GeV) & 45.6 & 80 & 120 & 182.5\\
\hline
Bunch population ($10^{11}$) & 2.53 &2.91 & 2.04& 2.64\\
\hline
Bunches per beam & 9600 & 880 & 248 & 36\\
\hline
RF frequency (MHz) & \multicolumn{3}{c}{400} & 400/800 \\
\hline
RF Voltage (GV) & 0.12 & 1.0 & 2.08 &  4.0/7.25 \\
\hline
Energy loss per turn (GeV) & 0.0391 & .37 & 1.869 & 10.0  \\
\hline
Longitudinal damping time (turns) & 1167 & 217 & 64.5 & 18.5\\
\hline
Momentum compaction factor $10^{-6}$ & \multicolumn{2}{c|}{28.5} & \multicolumn{2}{|c|}{7.33}  \\ 
\hline
Horizontal tune/IP & \multicolumn{2}{c|} {55.563} & \multicolumn{2}{|c|}{100.565} \\
\hline
Vertical tune/IP & \multicolumn{2}{c|} {55.600}  &  \multicolumn{2}{|c|}{98.595} \\
\hline
Synchrotron tune & 0.0370 & 0.0801 &  0.0328 & 0.0826\\
\hline
Horizontal emittance (nm) & 0.71 & 2.17 & 0.64 & 1.49  \\
\hline
Verical emittance (pm) & 1.42 &  4.34 & 1.29 &  2.98\\
\hline
IP number & \multicolumn{4}{|c|}{4} \\
\hline
Nominal bunch length (mm) (SR/BS)$^*$ & 4.37/14.5 & 3.55/8.01 & 3.34/6.0 & 2.02/2.95\\
\hline
Nominal energy spread (\%) (SR/BS)$^*$ & 0.039/0.130 & 0.069/0.154 & 0.103/0.185 & 0.157/0.229 \\
\hline
Piwinski angle (SR/BS)$^*$ & 6.35/21.1 & 2.56/5.78 & 3.62/6.50 & 0.79/1.15  \\
\hline
$\xi_x / \xi_y$ & 0.004/0.152 & 0.011/0.125 & 0.014/0.131 & 0.096/0.151 \\
\hline
Horizontal $\beta^*$ (m) & 0.15 & 0.2 & 0.3& 1.0 \\
\hline
Vertical $\beta^*$ (mm) & 0.8 & 1.0 & 1.0 & 1.6 \\
\hline
Luminosity/IP ($10^{34}$/cm$^{2}$s) & 181 & 17.4 & 7.8 & 1.25 \\
\hline
\multicolumn{1}{c}{$^*$SR: syncrotron radiation, BS: beamstrahlung} \\
\end{tabular}
\end{table}

A particular importance is given to the study of the combined effect of beam-beam interaction and beam coupling impedance, which has a drastic impact on the stability of colliding beams and on the achievable collider luminosity. The beam-beam interaction alone with the extreme FCC-ee parameters has already given rise to the several new important effects, such as beamstrahlung~\cite{telnov}, coherent X-Z instability~\cite{ohmi} and 3D flip-flop~\cite{shatilov}. The beam dynamics becomes much more complicated when also the beam impedance is taken into account~\cite{mikhail_c}, \cite{yuan}, \cite{fccee0},\cite{lin}.  

Without beam-beam interaction, the longitudinal impedance leads to bunch lengthening, synchrotron tune reduction and synchrotron tune spread increase, see~\cite{fccee1} for example. In addition, beyond the microwave instability threshold, the energy spread starts growing. In turn, the transverse impedance can trigger the well-known turbulent mode-coupling instability (TMCI) that can cause the beam loss~\cite{ipac21}. In combination with the beam-beam interaction, the longitudinal impedance results in variation of several beam parameters and, besides, it reinforces the X-Z coherent beam-beam instability. As it has been observed in numerical simulations for both CEPC~\cite{yuan} and FCC-ee~\cite{fccee0} the stable tune areas, free of instability, get narrower and their positions on the tune diagram are shifted because of the impedance related synchrotron tune reduction. A theoretical explanation of this effect can be found in~\cite{lin}.

It has been shown in~\cite{fccee0} that for the original CDR parameters the interplay between the beam-beam interaction, beamstrahlung and the impedance makes the stable tune areas available for the collider operation too small. So in order to overcome the problem, two mitigation techniques, that is the use of the harmonic cavities and an increase of the lattice momentum compaction factor, have been proposed. Following that work, a lattice with twice as higher momentum compaction factor with respect to CDR had been chosen for further studies.

Besides, the collider parameters are continuously evolving according to the lattice development and to beam dynamics studies. The major relevant change recently proposed is the possibility of using 4 interaction points (4 IPs) instead of 2 IPs, as requested by detector experts and particle physicists. Finally, also the FCC-ee impedance model is being constantly updated and refined closely following the vacuum chamber design. As a consequence, in order to be sure that these recent variations do not harm the collider performance, full scale 3D simulations of the beam-beam interaction with 4 IPs, with the inclusion of the updated longitudinal beam coupling impedance, should be performed. 

This paper, related to beam-beam and collective effects with the new parameter list, is organised as follows. In Sec.~\ref{sec:1} we discuss the updated impedance model, while its effects on the longitudinal beam dynamics are considered in Sec.~\ref{s:long_dyn}. While in all previous studies of the TMCI instability only the transverse resistive wall impedance was taken into account, Sec.~\ref{s:transv_dyn} shows the results of the transverse beam dynamics simulations using the more complete transverse impedance model. The TMCI threshold is evaluated including both transverse and longitudinal impedances, and the approach is useful to understand how the longitudinal beam dynamics affects this transverse instability. Finally, the 3D self-consistent simulations of the beam-beam interactions with 4 IPs, including the updated longitudinal impedance, are described in Sec.~\ref{s:beam-beam}. In order to mitigate the observed severe reduction of the stable tune areas in collision with 4 IPs, a horizontal betatron function reduction in the collision points is proposed in Sec.~\ref{s:mitigation}. In Conclusions the findings of these studies are summarised and plans for future studies are outlined.

\section{Updated impedance model}
\label{sec:1}

Since the design of the vacuum chamber and of the machine devices is still under development, the FCC-ee impedance model is constantly updated. In particular, in previous studies, such as~\cite{fccee0} and \cite{fccee2}, only the longitudinal impedance was included in the evaluation of collective effects. Just in~\cite{fccee1}, the transverse resistive wall impedance has been used for the study of the transverse mode coupling and coupled bunch instabilities.

The impedance revision that we are constantly carrying out has brought to a new transverse impedance model which includes now the same sources of the longitudinal one. Additionally, some elements, such as the bellows and the resistive wall of the vacuum chamber, which represent, so far, the main impedance sources, have been revised by considering more realistic models.

In the following subsections we first discuss these two revised models and then, at the end of the section, we present the total impedance that has been obtained in the three planes, longitudinal ($z$) and transverse $x$ and $y$. For the beam dynamics studies, only the dipolar contribution has been taken into account so far.

\subsection{Resistive wall}

In~\cite{fccee1} it was shown that the resistive wall impedance of a two-layer system with a thin coating has an additional imaginary term,  with respect to a single layer, proportional to the coating thickness. As a consequence, in order to reduce the resistive wall contribution, which represents the most important impedance source for FCC-ee, the NEG layer, foreseen for mitigating the electron cloud build-up in the positron machine and for pumping reasons in both rings, should be as thin as possible. 

The experimental activity discussed in~\cite{fccee2} has shown that a NEG thickness between 100 and 200 nm is a good compromise with respect to the balancing between the limitations of activation and the coupling impedance. So far, the beam dynamics studies used a thickness value of 100 nm (see, e.g.~\cite{fccee0}). However, according to the vacuum experts~\cite{kersevan}, in order to guarantee a uniform coating all along the beam pipe, a value of 150 nm should be considered. 

\begin{figure}
\centering
\includegraphics[width=0.48\textwidth]{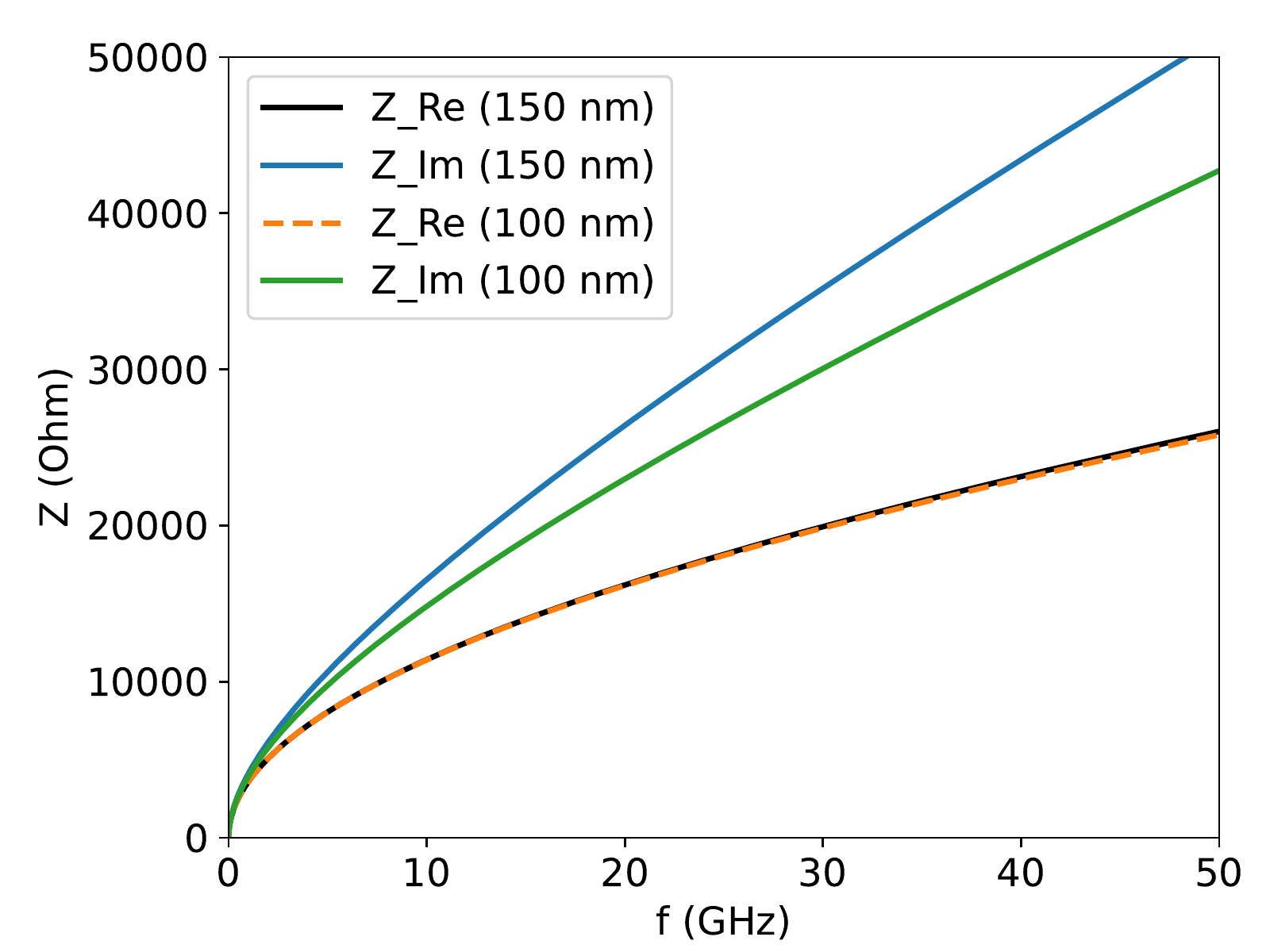}
\includegraphics[width=0.48\textwidth]{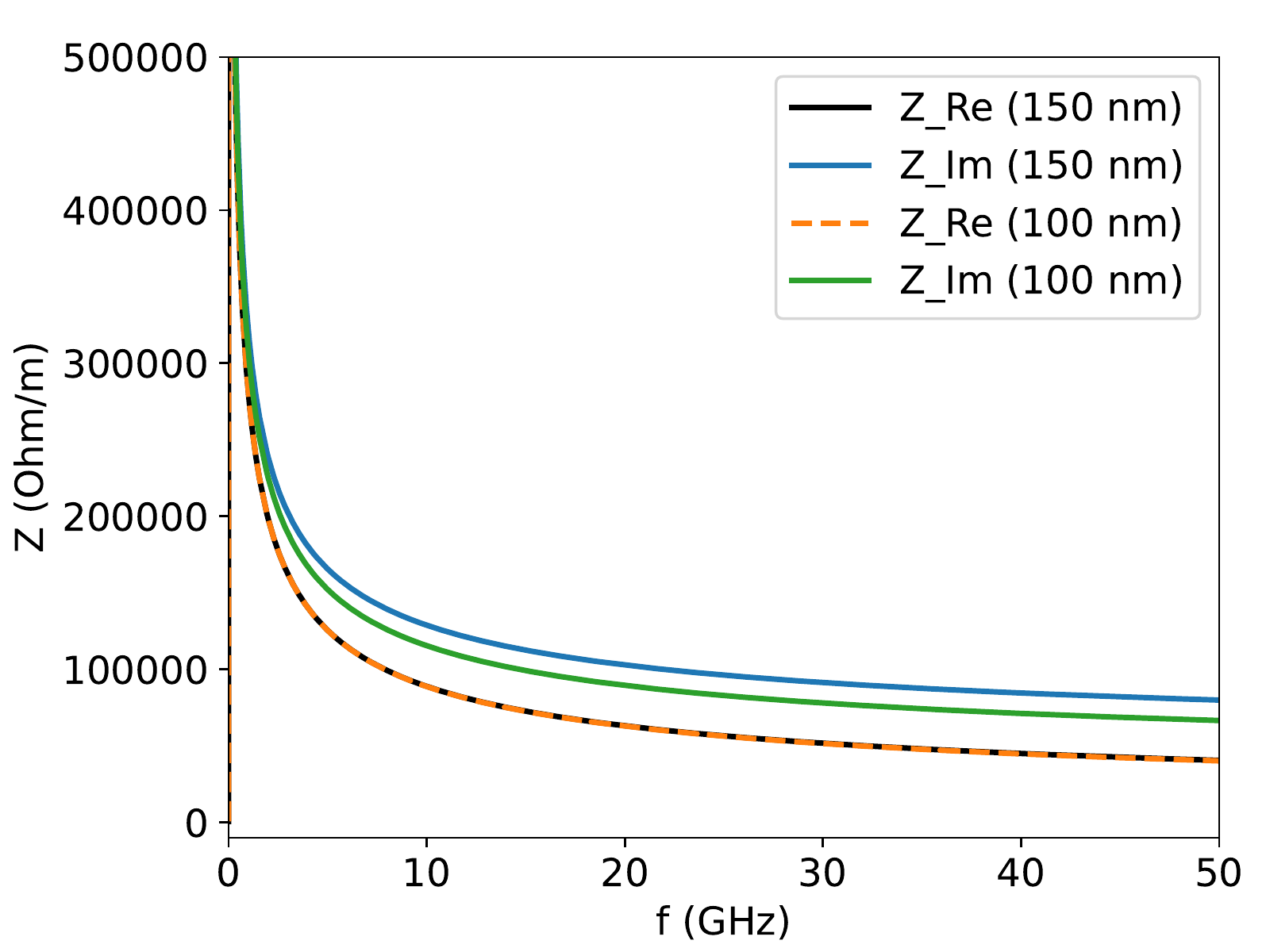}
        \caption{Longitudinal (left) and transverse (right) RW impedance for a circular pipe with two different NEG coating thicknesses: 100 and 150 nm.}
        \label{fig:RW_imp}
\end{figure}

The increase in the longitudinal and transverse impedance budget with respect to the 100 nm case is shown in Figs.~\ref{fig:RW_imp}, where the code IW2D~\cite{IW2D} has been used with four layers in circular geometry, as already discussed in ref.~\cite{fccee0}. As can be seen from the figure, and also according to Eqs.~(8) and (9) of~\cite{fccee1}, only the imaginary part of the impedance is affected by the increased thickness, and the loss factor does not change. Moreover it is possible to demonstrate that, in our frequency range of interest, only the first two layers (copper and NEG) are important.

The increased imaginary part of the impedance of course has an influence on beam instability thresholds, as it will be discussed in Sections~\ref{s:long_dyn} and \ref{s:transv_dyn}.

\begin{figure}
\centering
\includegraphics[width=0.5\textwidth]{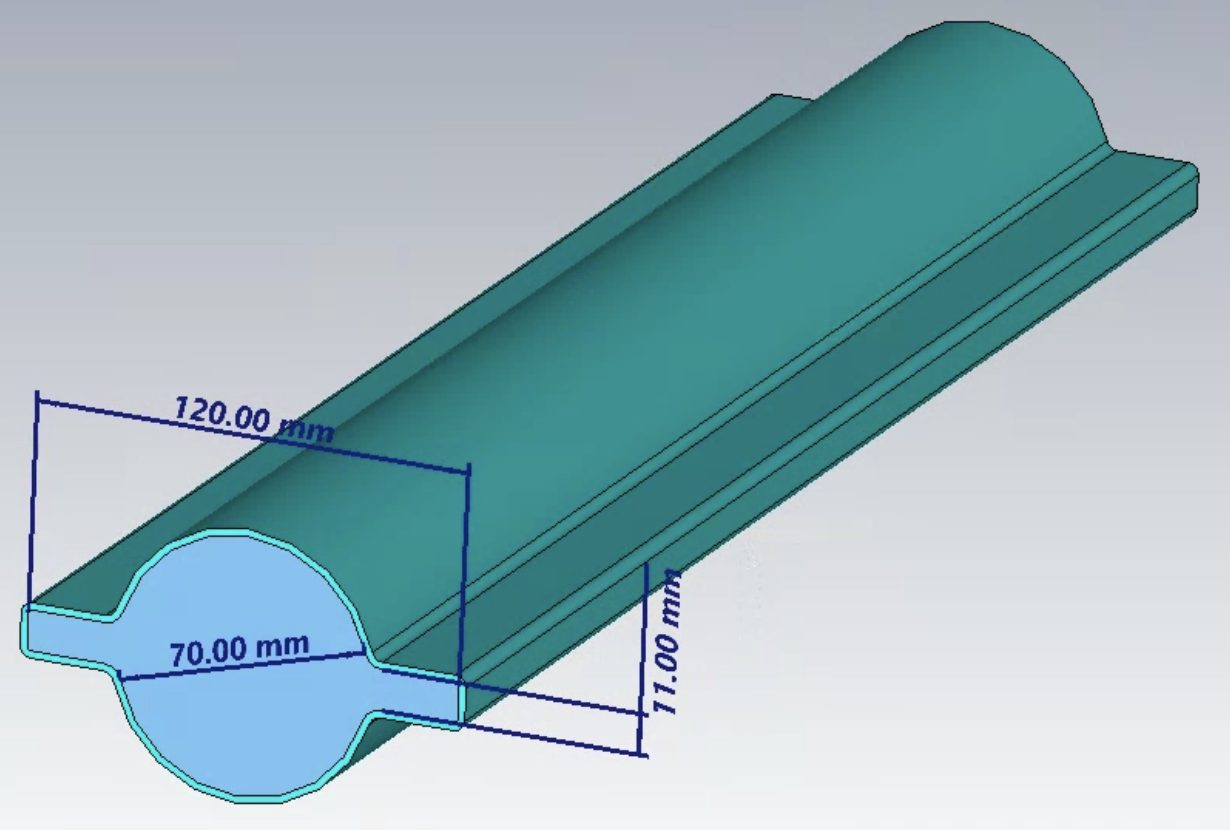}
        \caption{Beam pipe shape.}
        \label{fig:pipe}
\end{figure}

In addition to the increased coating thickness, we have also investigated the additional effect to the impedance due to the two lateral winglets of a more realistic vacuum chamber as shown in Fig.~\ref{fig:pipe}. The winglets are needed to place synchrotron radiation absorbers 'hidden' to the beam.

The figure has been obtained with CST Microwave Studio~\cite{CST}. Indeed the code IW2D, used so far, gives the impedance and wakefield only for circular and flat geometries. On the other hand, CST is not the most suited code to investigate the resistive wall impedance of a multi-layer system. For such a problem, by using the material type 'lossy metal', the code simulates the structure by using a one-dimensional surface impedance model. However, if the electromagnetic field penetrates through the layer, as is the case of the NEG coating foreseen in FCC-ee, one should use the 'normal type' material for each layer. This requires a very high number of mesh cells with too heavy computational resources.

In order to overcome this problem and have an estimate of the effects of the winglets, we have still used CST, but in an 'indirect' way as follows.

We have first evaluated the impedance of a single thick wall of a material with a relatively low conductivity (here we used $\sigma_c=10^5$ S/m). The use of this conductivity was necessary due to the fact that with copper and a reasonable length of the beam pipe (half a meter for mesh cells reasons), the impedance would have been too small and of the same order of the numerical noise. We observe, in any case, that for this chosen conductivity we are in a good conductor regime, characterised by $\sigma_c>> \omega \varepsilon_0$, with $\varepsilon_0$ the vacuum permittivity, up to frequencies much higher than those of our interest. Indeed we remember that, for a Gaussian bunch, the cut-off angular frequency is $\omega_c=c/(2\pi \sigma_z)$ with $c$ the speed of light and $\sigma_z$ the bunch length. For the minimum bunch length of FCC-ee, equal to 4.32 mm, this corresponds to a frequency of about 11 GHz.

Once the CST impedance of this model was determined, we divided it by its surface, one-layer, impedance $Z_{s1}$ and  multiplied it by the surface impedance of a double layer given by~\cite{elliptic} 
\begin{equation}
    Z_{s2}(\omega)=( 1+j)\sqrt{\frac{\omega Z_0}{2\sigma_c c}} \frac{\alpha \tanh \left[\frac{1+j}{\delta_1}\Delta \right] +1}{\alpha +\tanh \left[\frac{1+j}{\delta_1}\Delta \right] },
\end{equation}
where $\omega$ is the angular frequency, $Z_0$ the vacuum impedance, $\sigma_c$ the coating conductivity, $\delta_1$ the skin depth of the coating, $\Delta$ its corresponding thickness, and, for a good conductor, $\alpha \simeq \delta_1/\delta_2$, with $\delta_2$ the skin depth of the substrate, which is supposed to be of infinite thickness.

\begin{figure}
\centering
\includegraphics[width=0.48\textwidth]{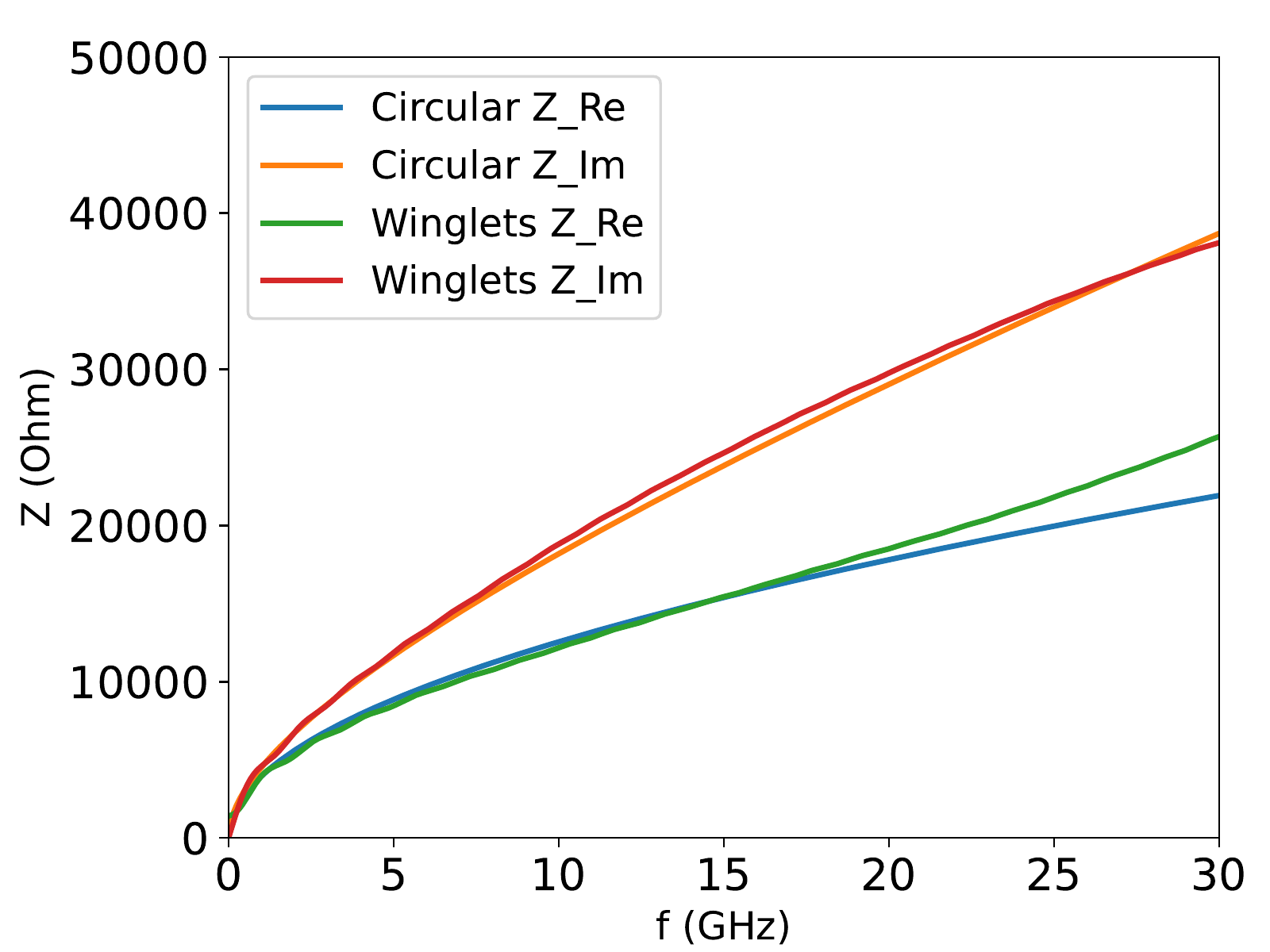}
\includegraphics[width=0.48\textwidth]{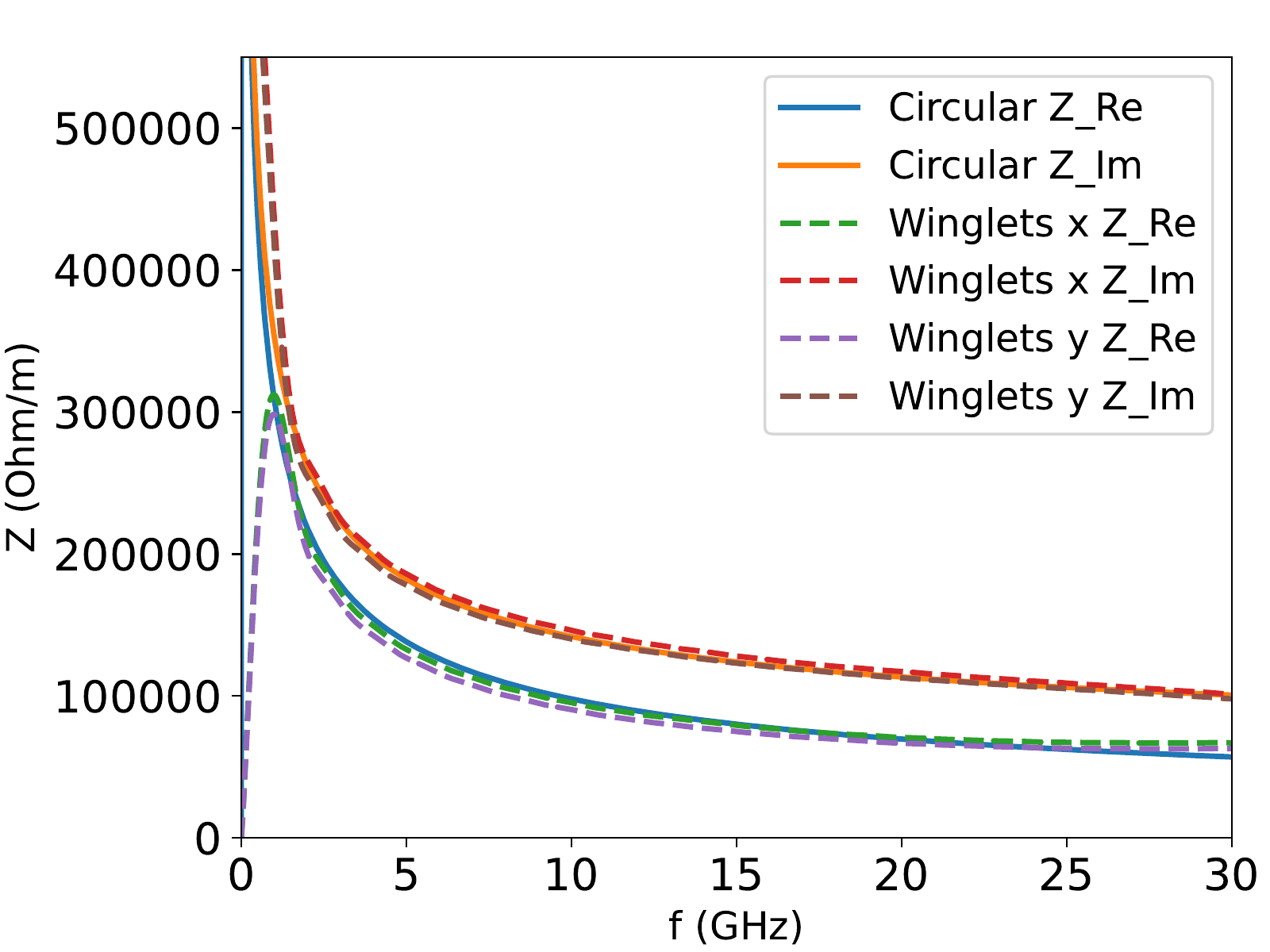}
        \caption{Resistive wall longitudinal (left) and transverse (right) impedance for FCC-ee obtained with CST by considering the winglets realistic model with a single infinite layer of material having a conductivity of $\sigma_c=10^5$ S/m re-scaled with the surface impedance of a double layer and compared with the results of IW2D with four layers for a circular pipe, and multiplied by a factor 1.1.}
        \label{fig:RW_winglets}
\end{figure}

The final impedances, for both the longitudinal and transverse planes, are represented in Figs.~\ref{fig:RW_winglets}, in the left-hand and right-hand side, respectively. As a check of the validity of this method, we have compared the results with those of IW2D of circular geometry, four layers, and multiplied by a form factor of 1.1 in both planes.

The results of CST have been obtained by considering a Gaussian bunch of 2 mm bunch length, thus having a cut-off frequency of about 24 GHz, and this is the reason why, in this comparison, we have arrived up to 30 GHz. In this frequency range there is a good agreement between the two results.

The slightly different behaviour of the real part of the transverse impedance of CST at small frequencies is due to the lossy metal material type that has been used. Indeed, the frequency dependent skin depth of the fields must be much smaller than the thickness of the metal solid around the beam pipe (that we have chosen in CST to be 2 mm), but at such low frequencies this hypothesis is not verified by CST.

However, since the goal of this comparison was not to check the frequency behaviour given by IW2D, but to understand possible corrections necessary in the results due to the winglets, from the results we can conclude that, as expected, the winglets produce a very small perturbation to the resistive wall impedance of the order of 10\%. We can take into account this effect by multiplying the impedance of the circular pipe given by IW2D by a factor 1.1. This factor can be seen as an approximated numerical 'Yokoya form factor'~\cite{yokoya} valid in the relativistic case and for the geometry with the winglets.

Finally, we observe that the winglets give also a contribution to the quadrupolar impedance due to the breaking of the cylindrical symmetry. However, by using the method just illustrated, we have evaluated a factor of about 20 less that that of the the dipolar impedance, and, in this paper, we neglect this quadrupolar contribution.

\subsection{Bellows}
\begin{figure}
\centering
\includegraphics[width=.3\linewidth, height=8cm]{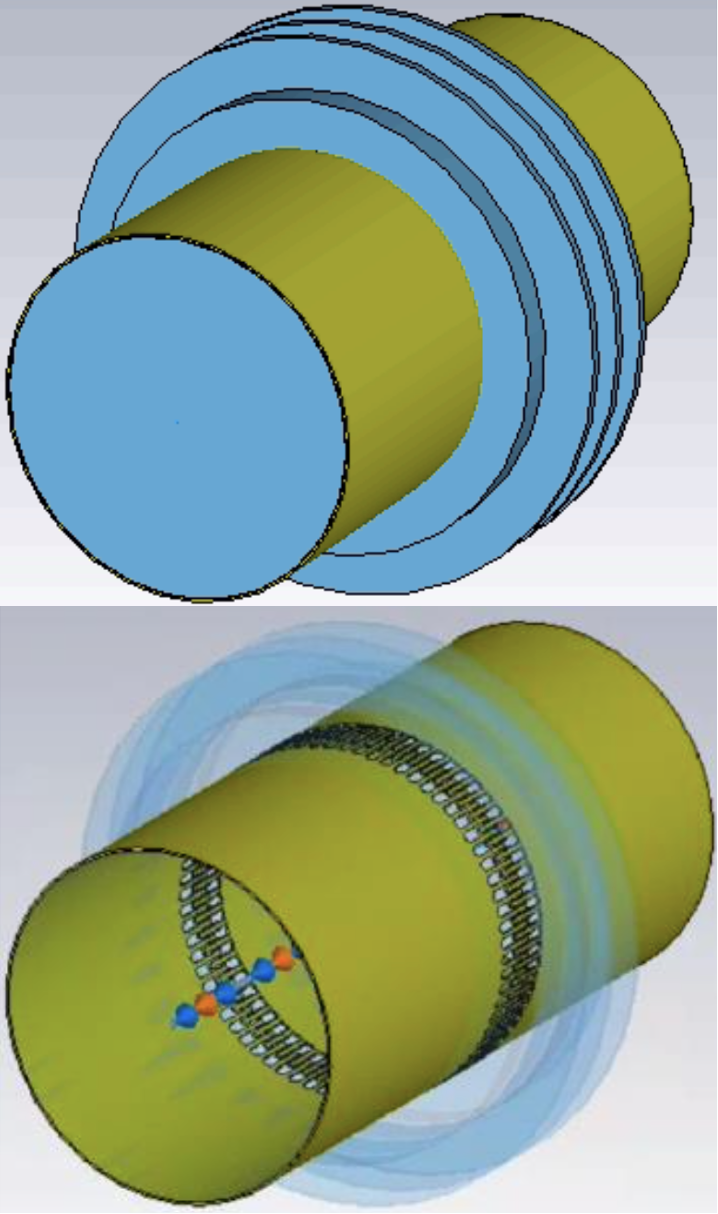}
\includegraphics[width=.3\linewidth, height=8cm]{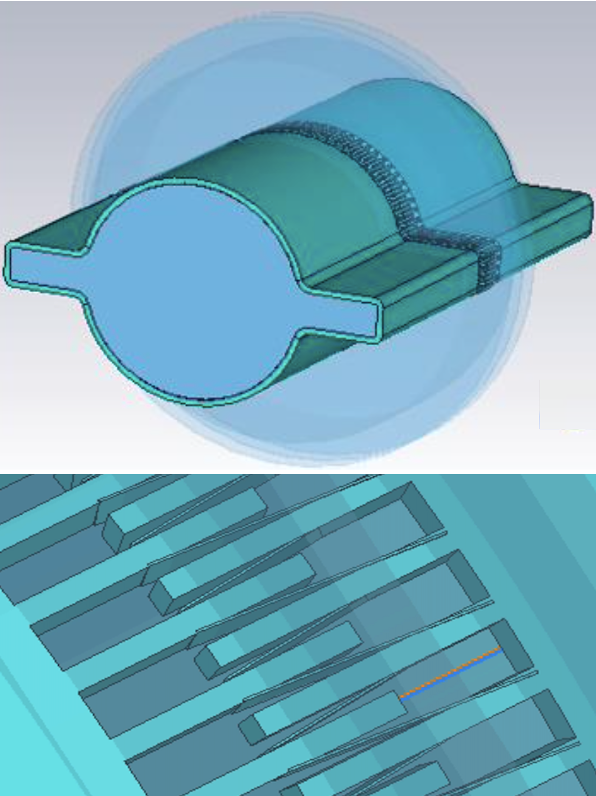}
\includegraphics[width=.3\linewidth, height=8cm]{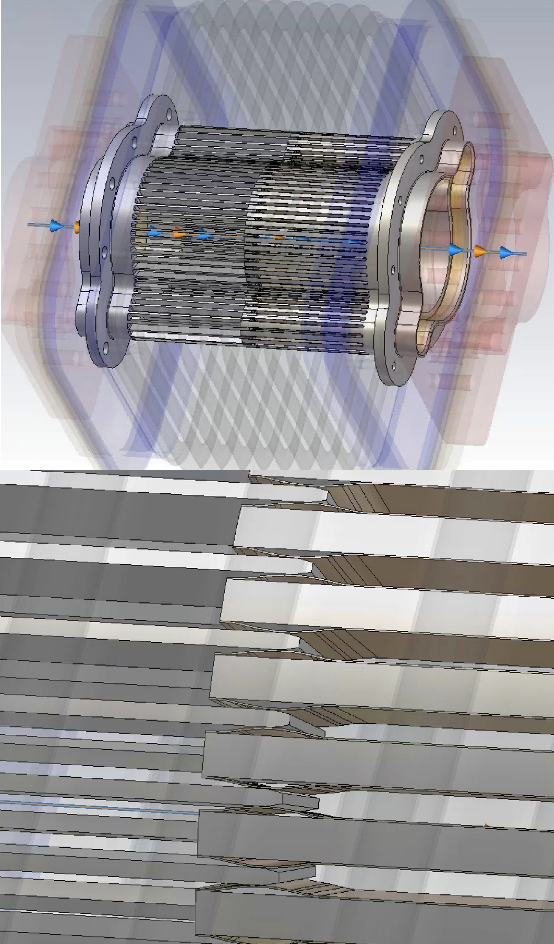}
\caption{Simulated models of FCC-ee beam vacuum chamber including bellows. Left: simplified model with circular geometry, centre: simplified model with winglets, right: realistic model.}
\label{fig:bellows}
\end{figure}

The presence of the winglets, on the other hand, plays an important role in the impedance of the bellows. Indeed, the early stage of the design featured a simplified model that considered the beam vacuum chamber with a circular profile without the winglets, as shown in Figs. \ref{fig:bellows}, left-hand side. The main contribution to the broadband impedance, both in the longitudinal and transverse planes, is due to the modes trapped between the beam pipe and the bellows, a space that, from the electromagnetic point of view, can be thought as a cavity coupled with the beam through the apertures due to the RF fingers. The presence of the winglets changes this space, thus modifying the parameters of the modes and their contribution to the broadband impedance.

The study of the bellows has been performed in three different steps, each one considering a design more realistic with respect to the previous one. The corresponding models are shown in Figs.~\ref{fig:bellows}, left (simplified circular shape), centre (circular shape with the addition of the winglets), and right-hand side (mechanical design provided by the CERN vacuum group~\cite{bellows}).

A crucial component of the device is the RF shielding with comb-type fingers and small electric fingers to ensure the electric contact between the two sides of the shielding. These fingers are shown in the lower part of the figure, at centre (for the first two models) and at the right-hand side (for the realistic model). The contribution of the shielding is fundamental to suppress the low frequency resonances due to the bellows which otherwise would lead to a high impedance contribution. 

The realistic model on the right side of the figure consists of a vacuum chamber designed with a mechanical CAD, with different comb-type fingers that best suite the bellow's shielding, and a realistic version of the bellows with squared profile, as shown in the top right part of the same figure. 

Preliminary simulations and most of the convergence studies related to the impedance behaviour have been carried out using the simplified CST models, firstly with the circular beam chamber and then including the winglets, since the realistic model required heavy computational resources. Afterwards, simulations on the realistic model were carried out to verify the results obtained with the simplified versions.

For all the considered models, longitudinal and transverse simulations have been performed with the wakefield solver of CST, and they were preliminary focused on the numerical convergence of the results, which turned out to be the most critical and challenging part for the electromagnetic characterisation of the device.

Actually, the complexity of the simulations deriving from the small mesh size, required to proper model the tiny fingers of the shielding, led to time consuming simulations and to the need of important computational resources. The situation was even more cumbersome in performing transverse simulations due to the limitation in the use of symmetries to reduce the number of mesh cells. In this scenario, the achievement of the numerical convergence turned out to be essential to allow the correct study of the problem under consideration and in many situations this required to use a large number of mesh cells, up to one billion.

\begin{figure}
\centering
\includegraphics[width=0.48\textwidth]{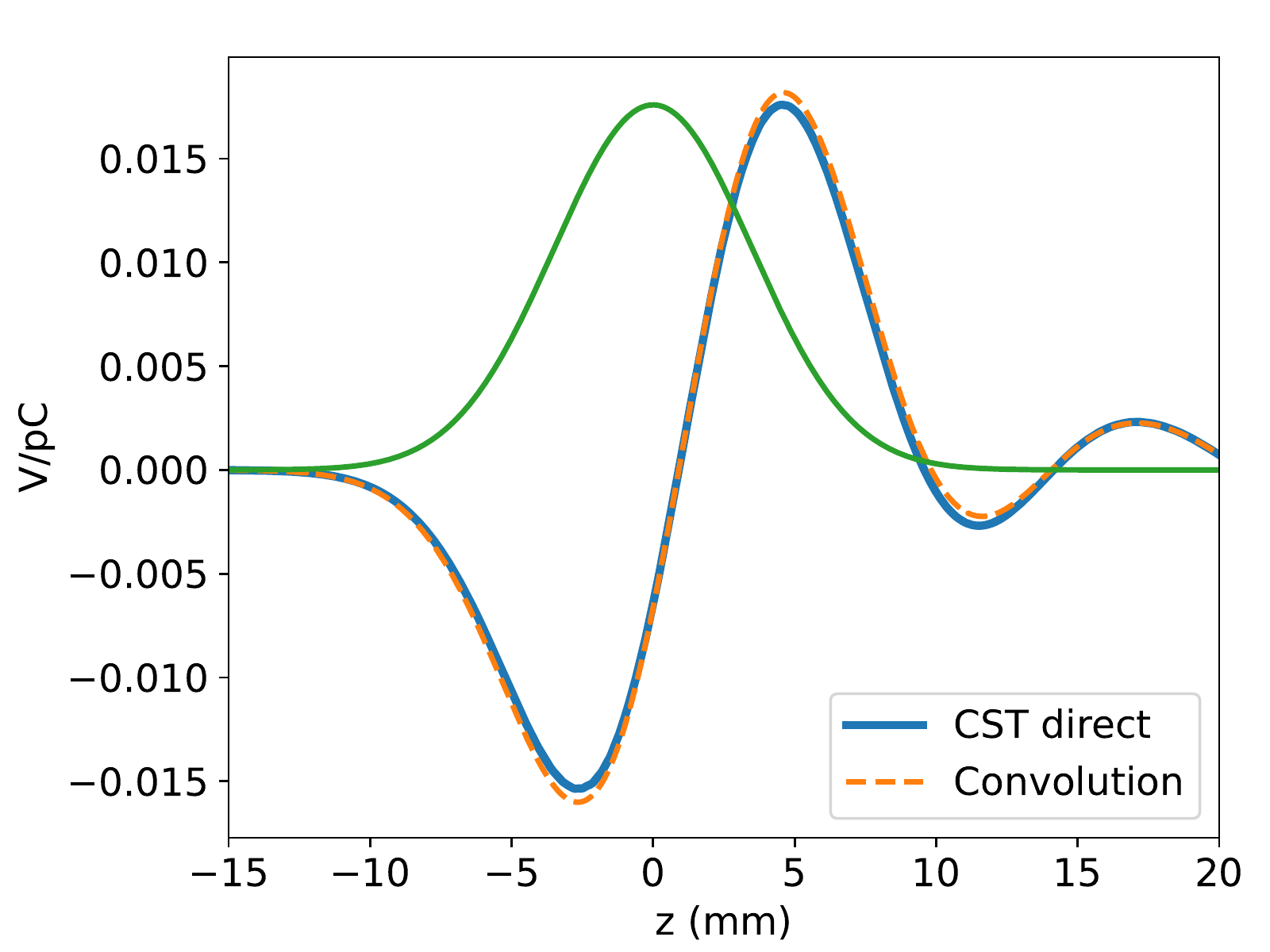}
\includegraphics[width=0.48\textwidth]{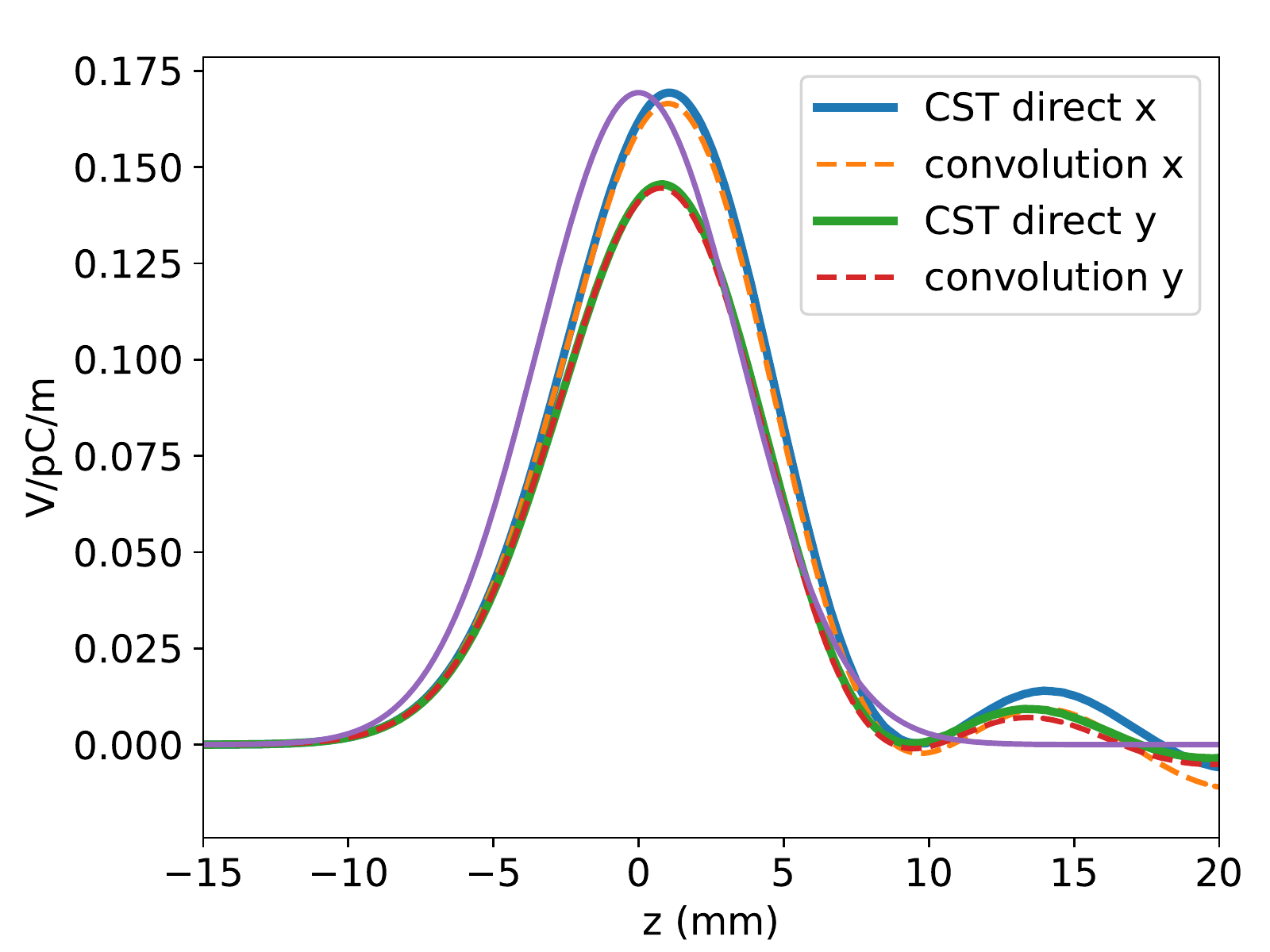}
\caption{Comparison of longitudinal (left) and transverse (right) wake potential of a 3.5 mm Gaussian bunch for a single bellow between the direct results of CST and the reconstructed wake by means of the convolution integral.}
\label{fig:bellows_comp}
\end{figure}

One check that was performed after the convergence studies, was the possibility to reconstruct the wake potentials of a longer Gaussian bunch, for example of 3.5 mm, with that of 0.4 mm used as pseudo-Green function. The comparison between the direct result of CST and the reconstruction by means of the convolution integral is shown in Figs.~\ref{fig:bellows_comp} for both the longitudinal (left-hand side) and transverse (right-hand side) planes.

\begin{figure}
\centering
\includegraphics[width=0.56\textwidth]{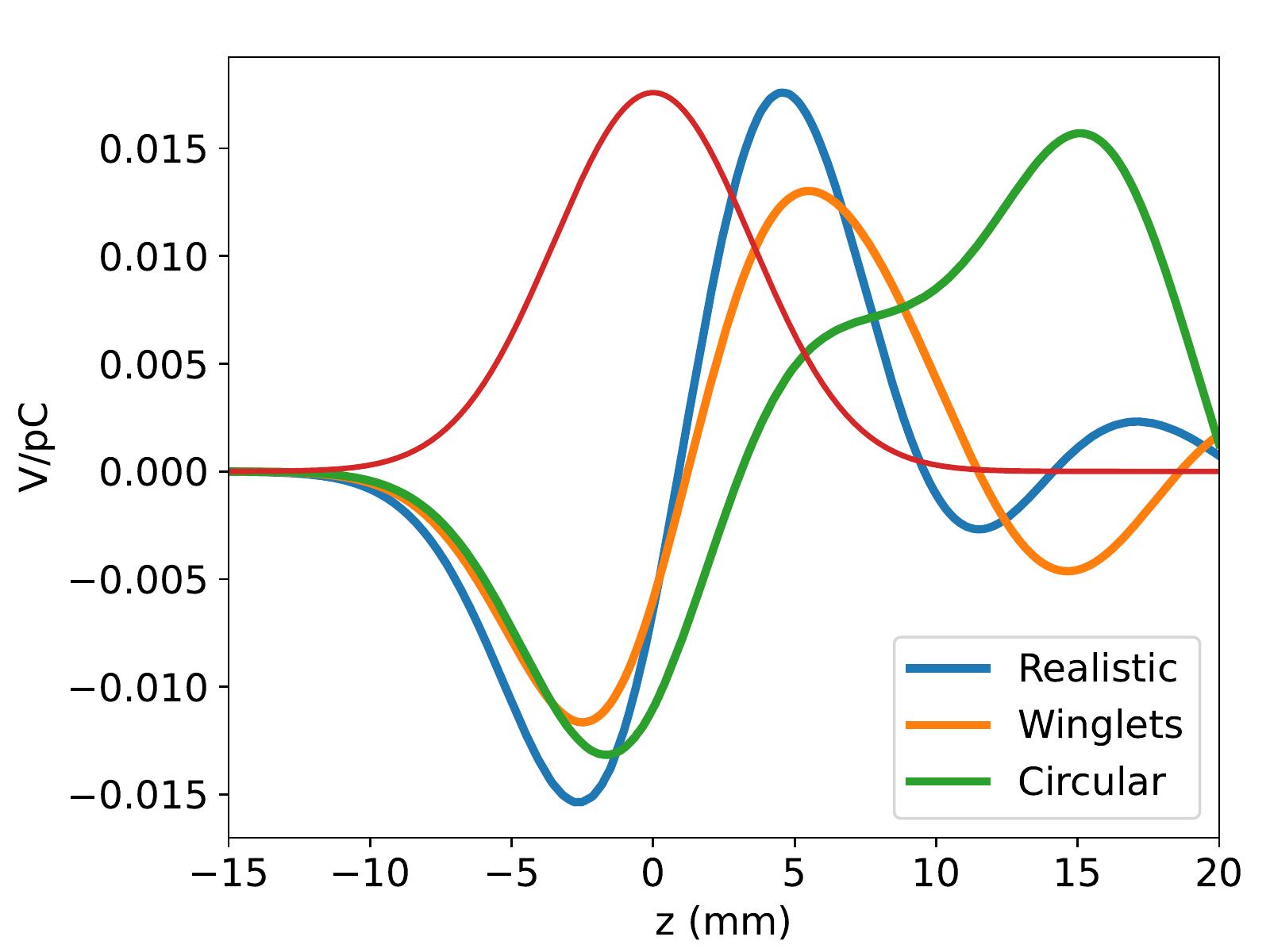}
\includegraphics[width=0.42\textwidth]{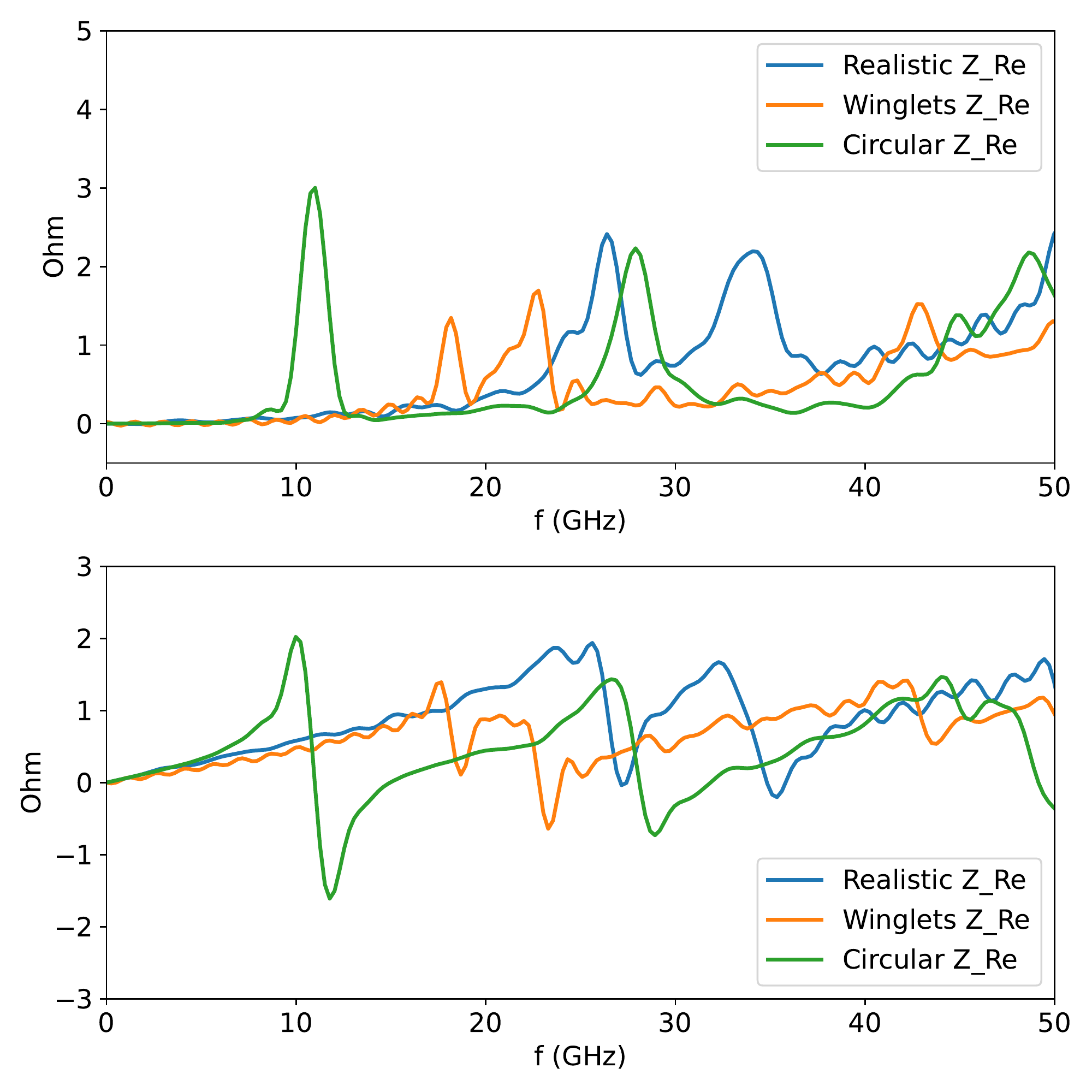}
\caption{Longitudinal wake potential and impedance for the three studied models of the bellow.}
\label{fig:Long}
\end{figure}

The results of the longitudinal wakefields for 3.5 mm Gaussian bunch and the impedances of all the three studied models by considering a single bellow are reported in Figs. \ref{fig:Long} for comparison in the left and right-hand side, respectively.

The most obvious aspect is the absence of the first resonance of the circular model around 11 GHz in both the other two models with the winglets. This behaviour has been carefully studied and it turned out that the first resonance is related to the coupling between the beam vacuum chamber and the cavity of the bellow. When the radius of the beam chamber approaches the radius of the bellow cavity, the resonance decreases its amplitude more and more until it is suppressed. According to these results, it is not surprising that the resonance is almost suppressed in the case of the chamber with winglets, since the horizontal aperture of the chamber approaches the radius of the bellow cavity. As a result, the realistic vacuum chamber has a better performance from the impedance point of view. 

In the transverse plane we can draw similar conclusions: for the realistic model the low frequency dipolar impedance is negligible, almost null and the first resonance appears above 20 GHz. A similar situation occurs for the quadrupolar term of the impedance.

It is important to remind that for the total wakefield and impedance of the bellows we have to multiply these results by their number. The exact number is still unknown, however, if we consider 2900 dipole arcs 24 m long with bellows every 8 m~\cite{kersevan1} plus 2900 quadrupoles/sextupoles arcs as in the CDR~\cite{fcc}, we have a total of 11600 bellows. In addition to these, we need to take into account the bellows for the RF system, injection system, collimation, etc. Overall, to be conservative, we have overestimated them by using a total number of 20000 bellows.

\subsection{Total impedance}

\begin{figure}
\centering
\includegraphics[width=0.42\textwidth]{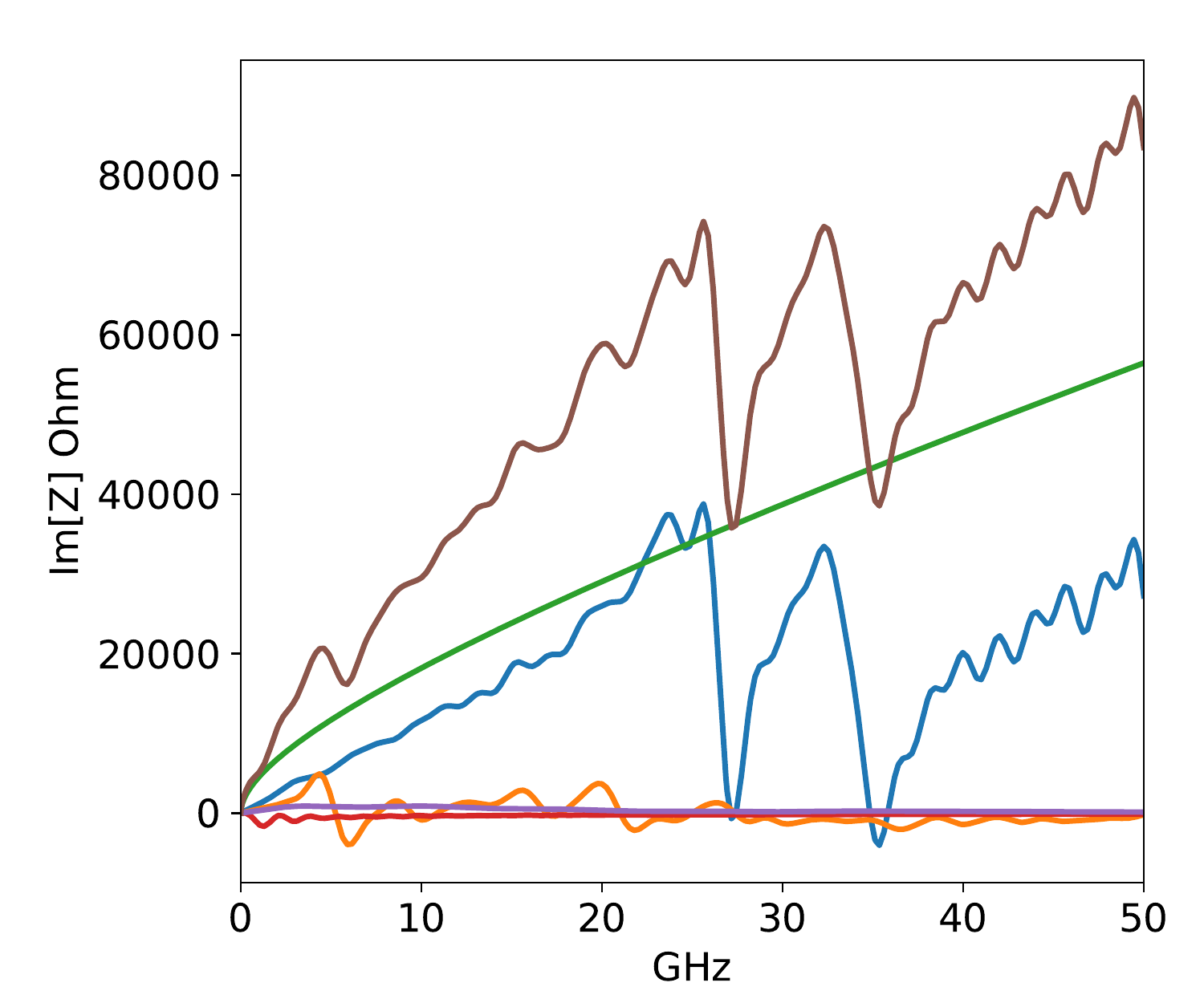}
\includegraphics[width=0.55\textwidth]{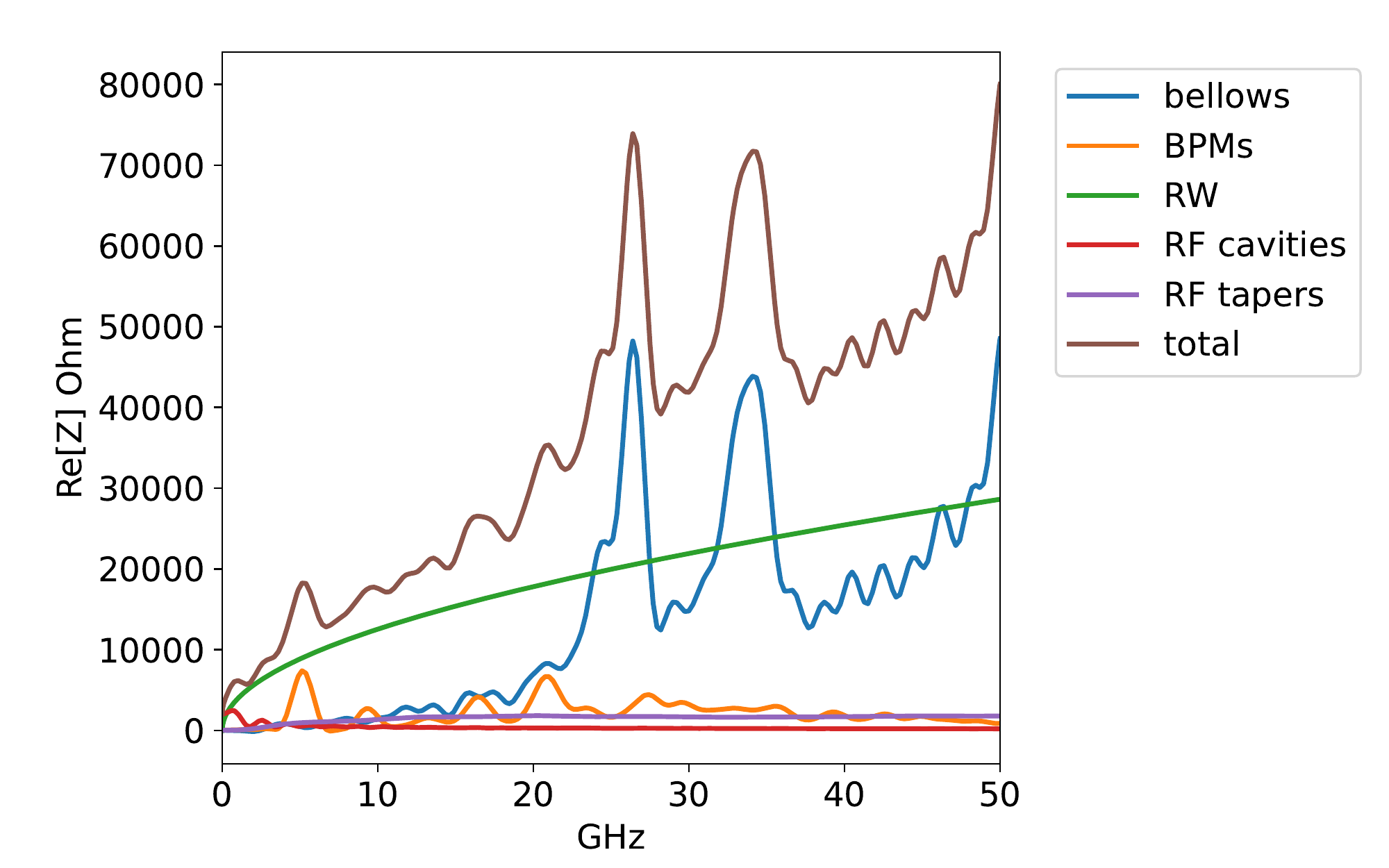}
\caption{Imaginary and Real part of the longitudinal impedance.}
\label{fig:Zlong_t}
\end{figure}

\begin{figure}
\centering
\includegraphics[width=0.42\textwidth]{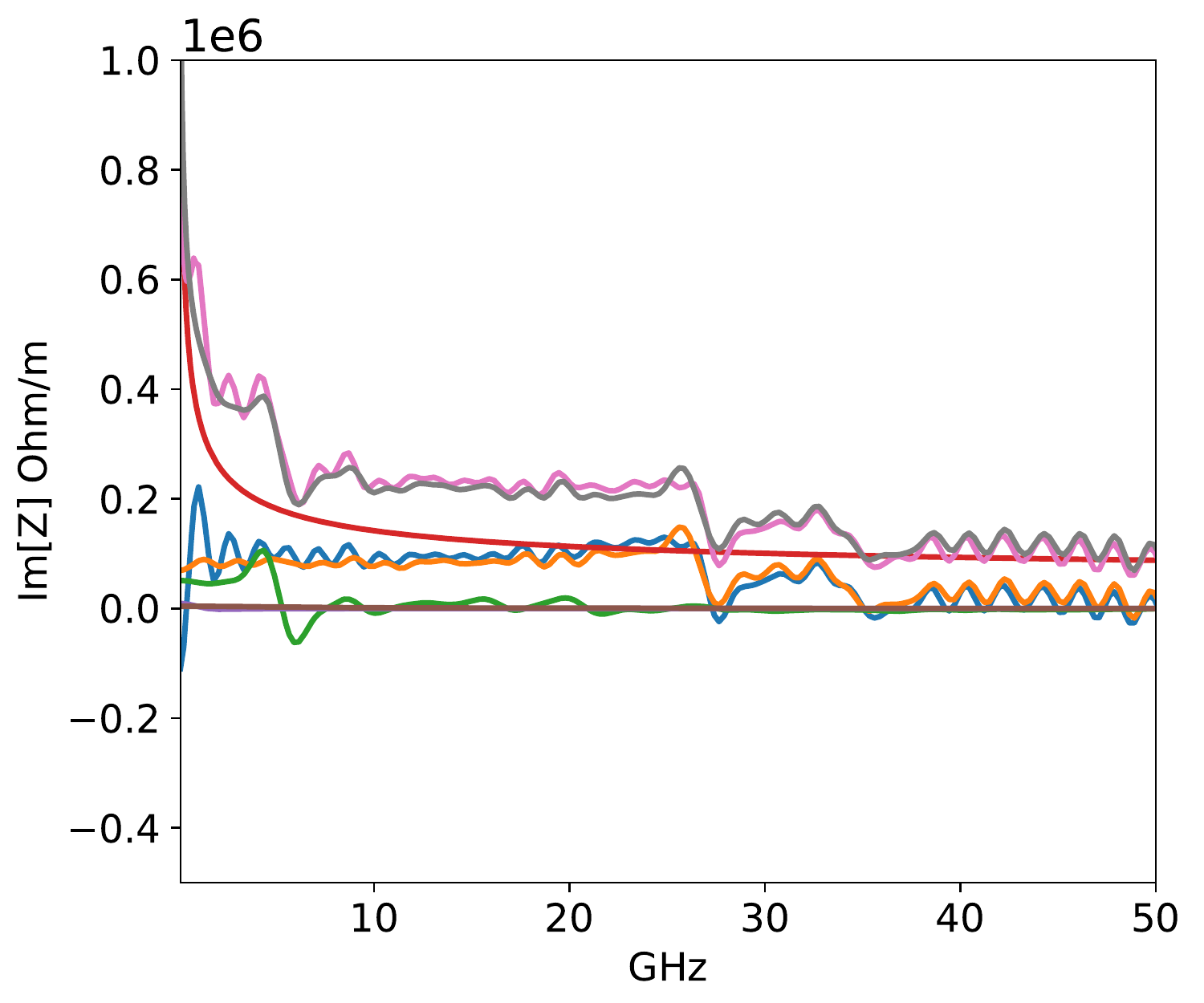}
\includegraphics[width=0.55\textwidth]{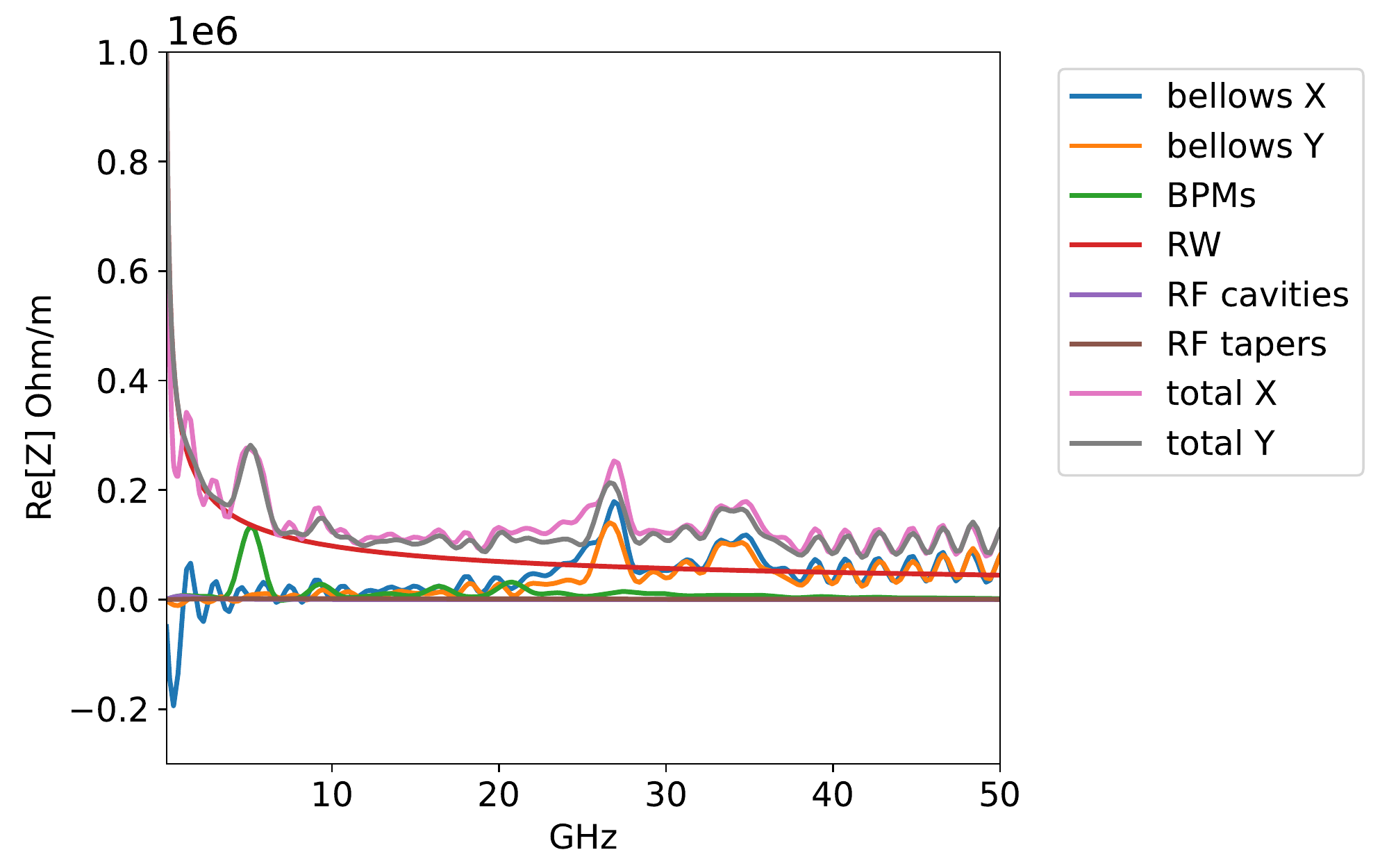}
\caption{Imaginary and Real part of the transverse impedance.}
\label{fig:Ztr_t}
\end{figure}

The impedance model evaluated so far takes into account also the beam position monitors and the RF system which includes the tapers connecting the cryo-modules. The total imaginary and real part of the broadband longitudinal and transverse impedances, together with the different contributions, are shown in Figs.~\ref{fig:Zlong_t} and Figs.~\ref{fig:Ztr_t}

It is important to note that, except for the resistive wall and the bellows, all the other devices give a small contribution (just the beam position monitors show a small peak around 5 GHz). Indeed, also the photon stoppers, or synchrotron radiation absorbers, have been evaluated in the past~\cite{fccee2}, \cite{ph_stop} resulting in a negligible contribution. A system that could be more critical, in particular in the transverse plane, is represented by the collimators. We are still missing this impedance and the study is in progress. We have to remind, however, that we almost doubled the impedance of the bellows in order to take into account, at this preliminary stage, other important and yet unknown possible impedance sources.

\begin{figure}
\centering
\includegraphics[width=0.6\textwidth]{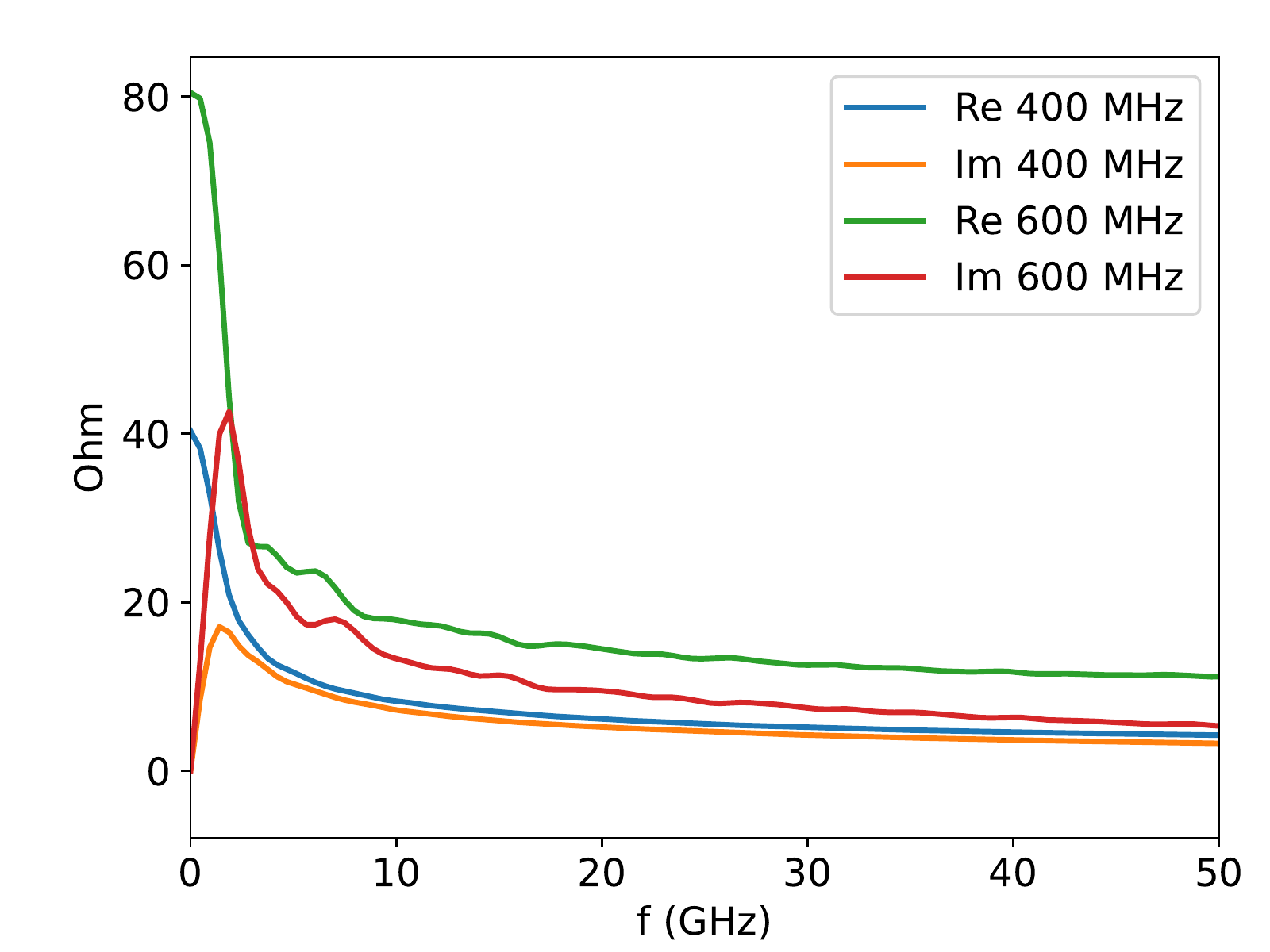}
\caption{Longitudinal impedance of 400 MHz single cell and 600 MHz double cell cavity.}
\label{fig:600MHz}
\end{figure}

In addition to the update of the resistive wall and bellows, other accelerator systems are also under development and may change in the future. For example, an alternative RF system is under study. It considers to use, instead of 52 single cell 400 MHz cavities grouped in 13 cryo-modules, each one having a double taper, as in the CDR~\cite{fcc}, a system using the so called two-cell 600 MHz Slotted Waveguide ELLipltical (SWELL) cavity~\cite{RF}. The longitudinal broadband impedance for a single cavity is shown in Fig.~\ref{fig:600MHz} in comparison with the 400 MHz one. 

\section{Longitudinal beam dynamics}
\label{s:long_dyn}

For the beam dynamics simulations by means of PyHEADTAIL code~\cite{pyht}, the results of which were preliminary compared with other tracking codes~\cite{SBSC1} \cite{SBSC2}, we have used, as pseudo-Green function, the wake potential of a very short Gaussian bunch, having a bunch length of 0.4 mm. This bunch length is smaller by a factor of 10 with respect to the minimum bunch length of Table~\ref{tab:1}.

The beam parameters for the low energy machine (Z pole) differ from those studied in ref.~\cite{fccee2}. Indeed we observe that here we have a higher single bunch population that gives stronger collective effects. On the other hand the momentum compaction factor is doubled and the nominal bunch length is increased, and these parameters mitigate the collective effects. Also the synchrotron tune is changed. As a conclusion, just from the analysis of the parameter list it is not easy to deduce the single bunch behaviour.

\begin{figure}
\centering
\includegraphics[width=0.45\textwidth]{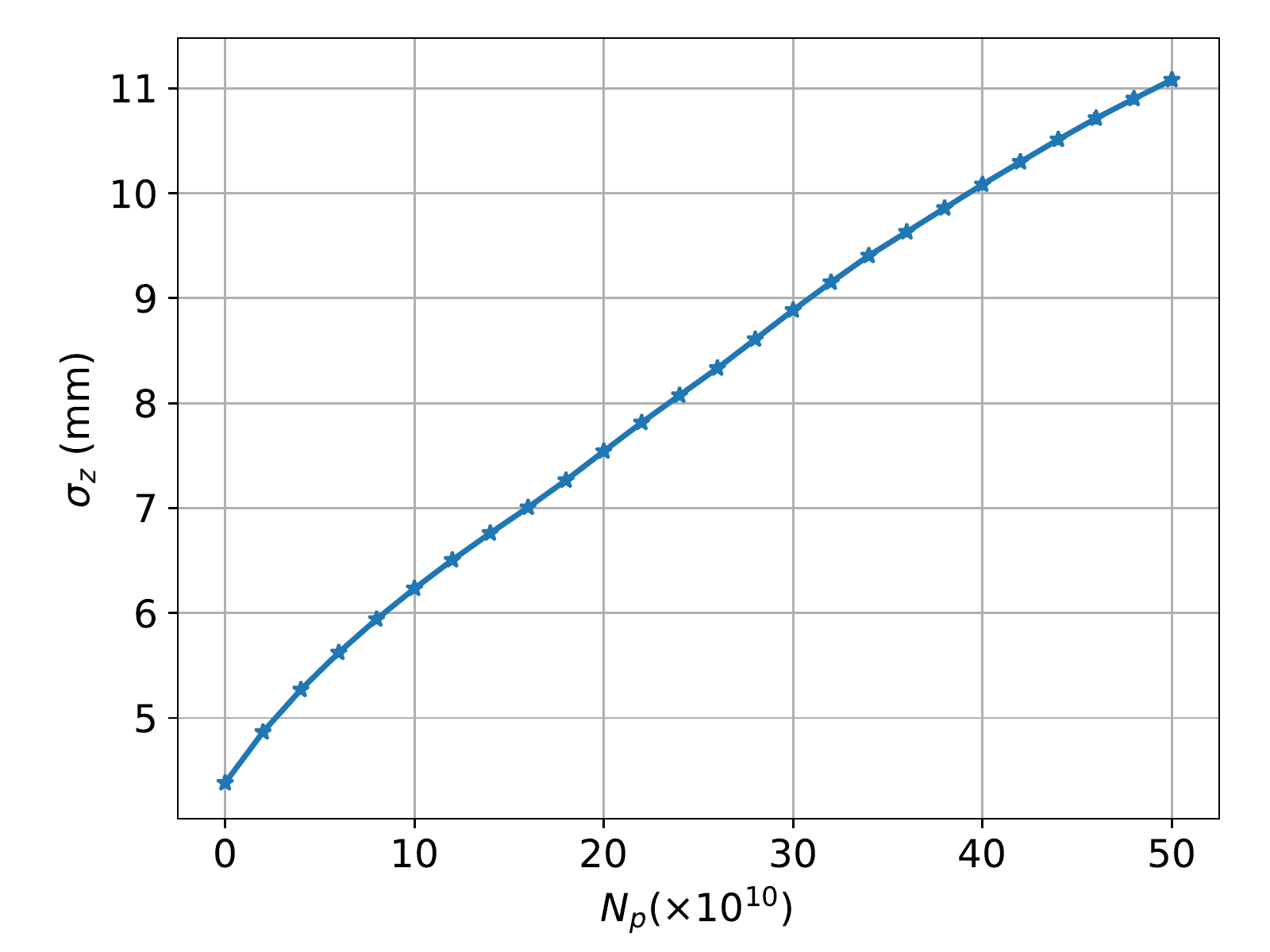}
\includegraphics[width=0.45\textwidth]{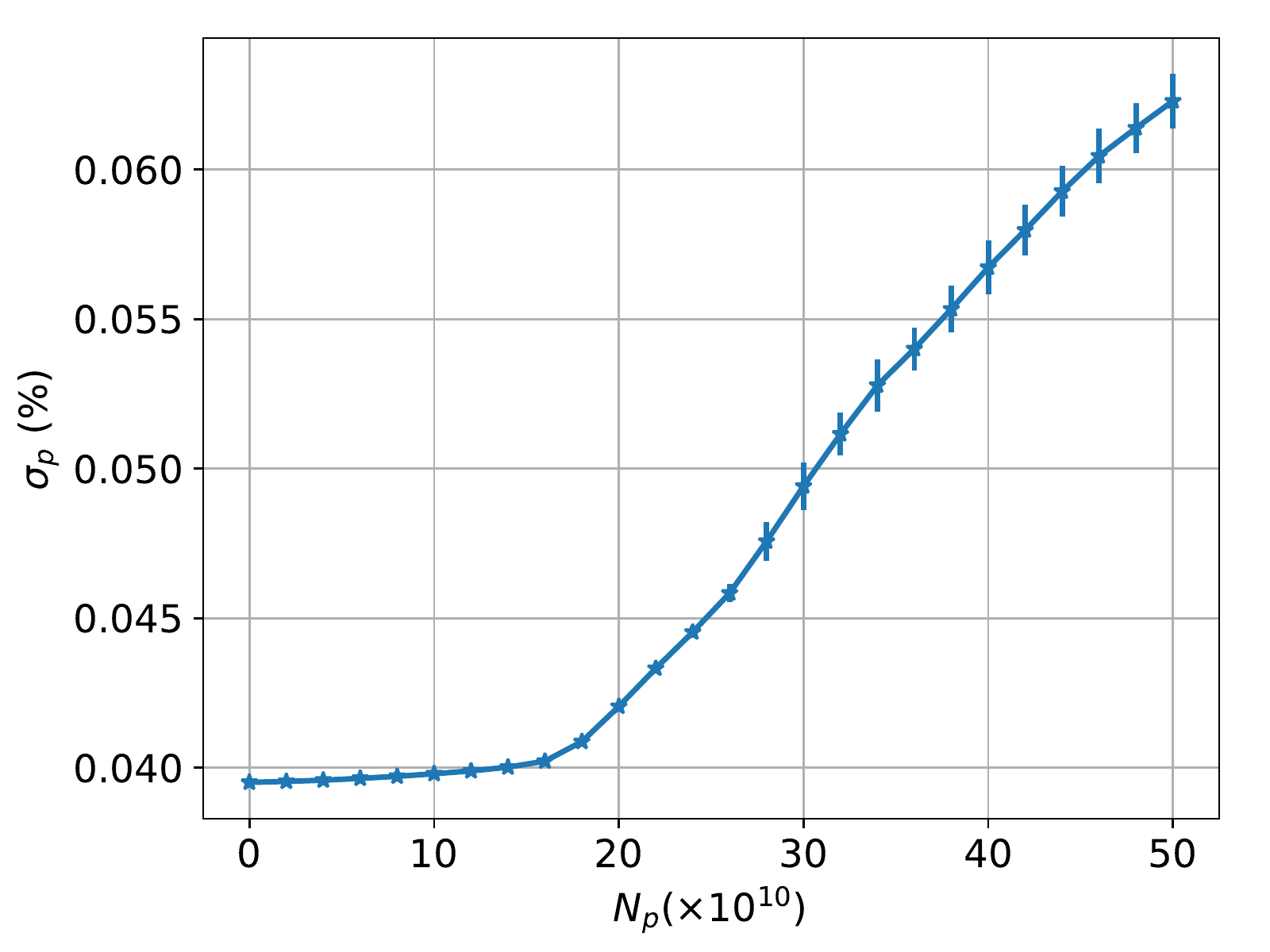}
\caption{Bunch length $\sigma_z$ (left) and energy spread $\sigma_p$ (right) as a function of the bunch population.}
\label{fig:bunchelenght}
\end{figure}

The rms bunch length $\sigma_z$ and energy spread $\sigma_p$ as a function of the bunch population are shown in the left-hand and right-hand side of Fig.~\ref{fig:bunchelenght}, respectively. 

If we compare these results with respect to those of ref.~\cite{fccee2} that refer to the CDR parameters, we observe that in the longitudinal plane the collective effects are less strong. The microwave instability threshold is around $2\times 10^{11}$, higher with respect to the previously studied case. 

For the nominal bunch intensity, the collider is expected to operate in a weak microwave instability regime. Indeed, as it can be seen in Fig.~\ref{fig:bunchelenght} for the nominal bunch population of $2.53\times10^{11}$, the energy spread growth is small, of the order of 10\%. Furthermore, the microwave instability should be suppressed in collision due to higher energy spread and the longer bunch as we have already observed in self-consistent simulations which include the contribution of the beam-beam interaction \cite{fccee0}. Figure \ref{fig:bunchelenght} can be useful to evaluate the bunch length and the energy spread during the collider commissioning phase without beam-beam collisions.

\section{Transverse beam dynamics}
\label{s:transv_dyn}

Previous studies of the FCC-ee transverse single bunch dynamics with inclusion of collective effects~\cite{ipac21} have used, as transverse impedance model, the resistive wall term. Here, for the first time, we performed the studies by taking into account the same sources as those of the longitudinal plane. 

As reference code for the simulations we still use PyHEADTAIL, but, in this case, preliminary comparisons were performed not with other tracking codes, as for the longitudinal plane, but with the Vlasov solver DELPHI~\cite{mounet}. Some results of these checks can be found in \cite{ipac21}. 

The main collective effect that we have studied in the transverse plane is due to the so called transverse mode coupling instability (TMCI) or strong head tail instability~\cite{chao}, ~\cite{ng}. In general the transverse motion of a bunch can be decomposed as a sum of coherent modes of oscillation, called eigenmodes, the coherent frequencies of which depend on the wakefields and, as consequence, on beam current intensity. At zero current, azimuthal modes are spaced by the synchrotron frequency. For FCC-ee it has been observed that, due to the transverse impedance, by increasing the current, the azimuthal mode '0' shifts towards the mode '-1' and, when they couple together, an exponential growth of the transverse emittance, with a rise time depending on the single bunch current, is observed.

\begin{figure}
\centering
\includegraphics[width=0.45\textwidth]{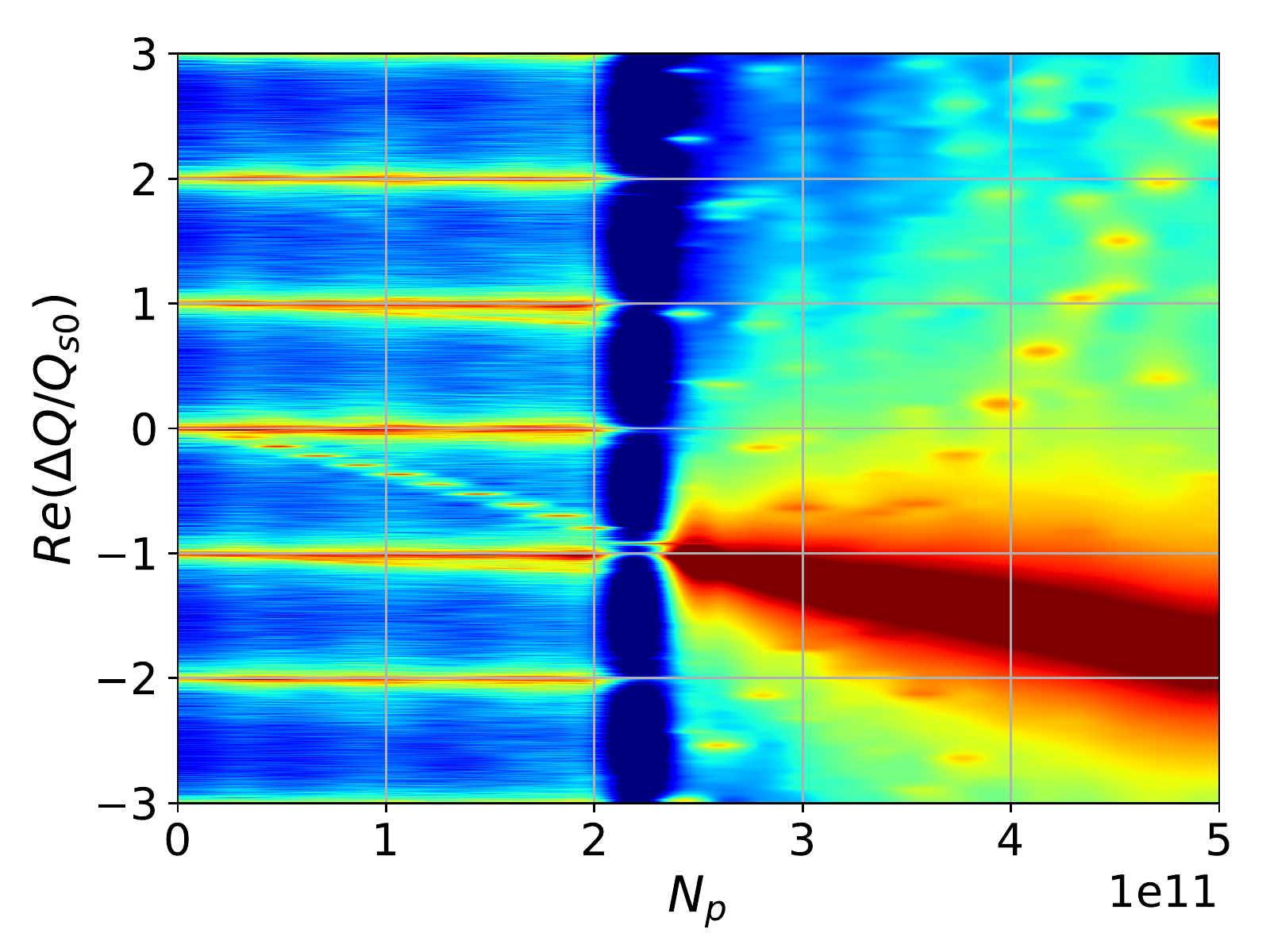}
\includegraphics[width=0.45\textwidth]{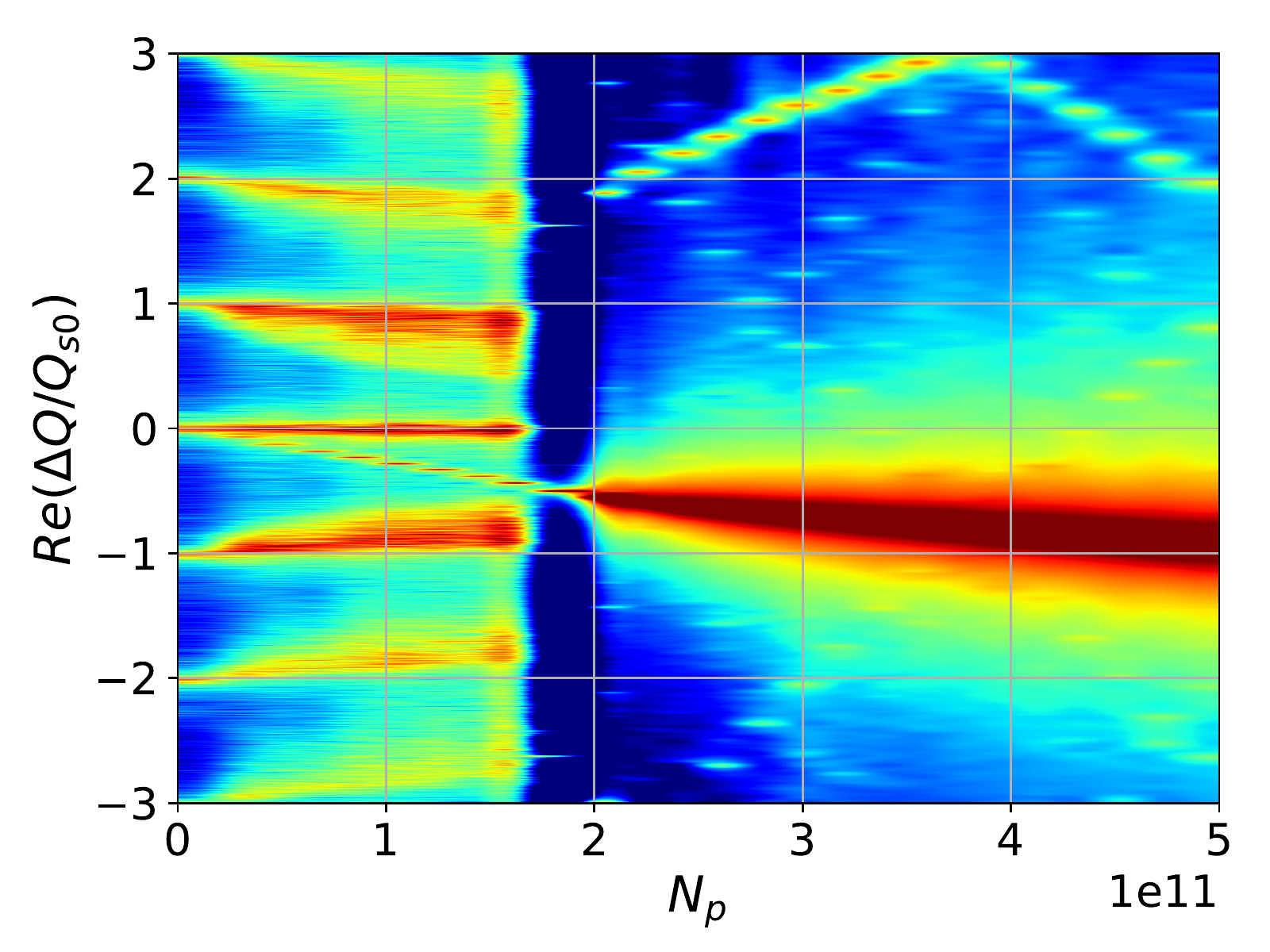}
\caption{Real part of the tune shift of the first azimuthal transverse coherent oscillation modes normalised by the synchrotron tune $Q_{s0}$ as a function of bunch population without (left) and with (right) the additional effect of the longitudinal wakefield.}
\label{fig:tmci}
\end{figure}

Coherent frequencies of the transverse modes can be derived from PyHEADTAIL results by using the method described in~\cite{mode_analysis}. In Fig.~\ref{fig:tmci}, left-hand side, we show the results of this analysis. The real part of the tune shift (with respect to the unperturbed betatron tune) of the first azimuthal transverse coherent oscillation modes, normalised by the unperturbed synchrotron tune $Q_{s0}$, and for the horizontal plane, is shown as a function of the bunch population. In the vertical plane, not reported here, the results are similar to the horizontal ones, with an instability threshold slightly different.

From the figure we can see that the mode coupling instability occurs at a bunch population lower than the nominal one of $N_p = 2.53\times 10^{11}$. In this case, differently from the longitudinal plane, the TMCI instability can be harmful since it is typically very fast in the electron-positron colliders and it can be hardly suppressed in collision. Therefore, the beam-beam simulations also including the transverse impedance is absolutely necessary and will be considered as a priority in our future studies. Nevertheless, we can expect that the beamstrahlung will play a beneficial role also in this case due to the longer bunches.  

The contemporary contribution of the longitudinal and transverse wakefields in a self-consistent way (with the exclusion of the beam-beam effects) can also be studied with PyHEADTAIL, and the results are shown in the right-hand side of Fig.~\ref{fig:tmci}, where, similarly to the left-hand side, we have reported the real part of the tune shift of the first azimuthal transverse coherent oscillation modes normalised by the synchrotron tune $Q_{s0}$ as a function of bunch population for the horizontal plane. By comparing this figure with the corresponding left-hand side, we observe that the longitudinal wakefield reduces the TMCI threshold. The same effect occurs in the vertical plane. 

Indeed the longitudinal wakefield gives a double contribution: from one side it increases the bunch length, thus reducing the strength of the transverse collective effects, but on the other side, it produces a synchronous tune shift and spread, in particular of the mode '-1' towards the mode '0', as can be seen in the right-hand side of the figure compared to that of the left-hand side, so  that the TMCI results with a lower threshold. As a consequence, TMCI in both horizontal and vertical planes occurs at a threshold lower than the nominal bunch population in the single beam case (no collisions).

It is important to underline that, while the microwave instability results in an oscillation of the bunch and an increase of the longitudinal emittance without any losses, the TMCI is more dangerous since it makes the bunch to be lost. As a consequence, it could not be possible to run the machine at the nominal intensity without colliding beams if no countermeasures are taken. Actually, the machine, in normal operation, is supposed to run in collision and the beamstrahlung helps in suppressing these instabilities.

However two important facts must be kept in mind: first, it could be necessary to run the machine with a single beam, as, for example, during the machine commissioning. In this case, if no mitigation tools will be devised, we must be aware that it will not be possible to commission the machine with the nominal bunch intensity. The second, important point is that, for FCC-ee, collective effects must be evaluated by considering the beamstrahlung in a self consistent way with the wakefields.

A final remark before concluding this section is related to the study that we have performed on localised or distributed wakefield kicks. So far all the simulations have considered the 'one turn map' determined by a transport of the particle coordinates along the machine plus the wakefield effect acting in one single point of the map. Since FCC-ee has a circumference of about 100 km, and the resistive wall and bellows are distributed all along the machine, we considered the question of possible differences between localised or distributed wakefield. PyHEADTAIL allows to split the 'one turn map' into different segments in the transverse plane but not longitudinally. We considered 59 segments for one turn, and, for each one, added the effect of the transverse wakefield: no substantial differences were noted. For the longitudinal plane some simulations with a machine divided into segments were performed some years ago with the code SBSC~\cite{SBSC1}, and, also in that case, no remarkable difference in the wakefield effect was observed.

\section{Beam-beam with 4 IPs}
\label{s:beam-beam}

As already discussed in detail in~\cite{fccee0}, the beam-beam interaction in FCC-ee contains a combination of several extreme beam parameters, such as very small emittances, small beta functions in the interaction points, a large Piwinski angle combined with the crab waist collision scheme~\cite{mikhail_a}, \cite{mikhail_b}, and very high intensity, that give rise to several new important effects, such as beamstrahlung~\cite{telnov}, coherent X-Z instability~\cite{ohmi} and 3D flip-flop~\cite{shatilov}. In turn, since the luminosity and both vertical and horizontal beam-beam tune shifts depend on the bunch length, taking into account the longitudinal impedance makes the beam dynamics in FCC-ee even more complicated~\cite{fccee0}, \cite{mikhail_c}, ~\cite{mikhail_d}, ~\cite{IBB}. First of all, this results in a reduction of stable working point areas available for the collider operation. In this section we study the effects of the updated longitudinal impedance on the collider performance for the new set of parameters with 4 IPs (see Table~\ref{tab:1}).

It is worth mentioning that the coherent horizontal-longitudinal (X-Z) instability represents one of the most critical phenomena for reaching the design collider performance. It is produced by the beam–beam interaction with large Piwinski angle and, differently from the coupling impedance induced collective instability, it is excited by the localized horizontal beam-beam force in the IPs. The instability generates a transverse beam size blow-up that can severely limit the stable tunes areas where the design luminosity can be achieved.

According to the semi-analytical scaling law of~\cite{ohmi}, the threshold of this instability is proportional to
\begin{equation}
\label{eq:1}
    N_{th} \propto \frac{\alpha_c \sigma_p \sigma_z}{\beta_x^*} \propto \frac{\nu_s} {\xi_x}.
\end{equation}

As it can be seen, the intensity threshold $N_{th}$ is proportional to the bunch length $\sigma_z$, energy spread $\sigma_p$ and the momentum compaction factor $\alpha_c$, while it scales inversely with the horizontal betratron function at the interaction point $\beta_x^*$.  Writing this scaling law in a different way, we find that the intensity threshold scales linearly with the synchrotron tune $\nu_s$ and it is inversely proportional to the horizontal beam-beam tune shift $\xi_x$. Such a scaling has a clear physical meaning considering that the coherent beam-beam synchro-betatron resonances are separated by the synchrotron tune while the strength of the resonances is proportional to the horizontal tune shift. 

For the case of 4 IPs placed symmetrically in the collider rings and having the same parameters at each collision point, it is possible to simplify the analysis considering only one interaction point in a collider which is 4 times shorter, having 4 times smaller synchrotron tune and 4 times longer damping times in terms of the revolution turns. Since the equivalent synchrotron tune is smaller and the damping time is longer, from Eq.~(\ref{eq:1}) one can expect that the X-Z instability will be more harmful in case of the 4 IPs in comparison with 1 IP or 2 IPs colliders.

Note that the scaling law~(\ref{eq:1}) is obtained without taking into account the longitudinal wakefields which affect both the bunch length and the energy spread. Besides, the wakefields produce a synchrotron tune reduction and a synchrotron tune spread. So, in order to obtain the correct results, full 3D self-consistent beam-beam simulations should be performed including the wakefields. These studies have been carried out by means of the simulation code IBB~\cite{IBB}.

\begin{figure}
\centering
\includegraphics[width=0.7\textwidth]{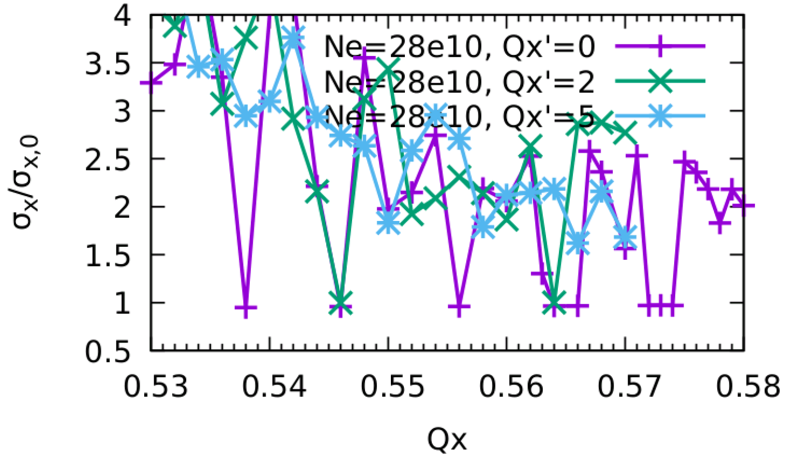}
\caption{Normalised horizontal beam size $\sigma_x/\sigma_{x,0}$ as a function of the horizontal tune for a bunch population of $N_p=2.8 \times 10^{11}$ at different chromaticities.}
\label{fig:X_Z_beta0_15}
\end{figure}

A first series of simulations was performed by considering a bunch population of $N_p=2.8 \times 10^{11}$. As it can be seen in Fig.~\ref{fig:X_Z_beta0_15}, where we have reported the normalised horizontal beam size $\sigma_x/\sigma_{x,0}$ as a function of the horizontal tune per IP, $Q_x$, the magenta line shows that it is not possible to find a sufficiently large tune area to operate the machine in stable condition in the tune range between 0.53 and 0.58. This is consistent with the above qualitative considerations.

As a first attempt to mitigate the instability, we considered the possibility of working with a positive chromaticity $Q_x'$ defined as
\begin{equation}
    Q_x' = \frac{\Delta Q_x}{\Delta p/p_0},
\end{equation}
with $p_0$ the nominal particle momentum. 

The chromaticity introduces a spread in the betatron tune that is expected to help in suppressing the instability, even if it reduces a bit the dynamic aperture. However, simulations by using $Q'_x =2$ and $Q'_x=5$ do not show any improvements, as reported in Fig.~\ref{fig:X_Z_beta0_15} with the green and cyan lines, respectively.

Since the bunch population used in the simulations is a bit higher than the nominal one foreseen in the parameter list of Table~\ref{tab:1}, we also tried to see if, by reducing the single bunch intensity, it was possible to find a sufficiently large stable tune region.

\begin{figure}
\centering
\includegraphics[width=0.7\textwidth]{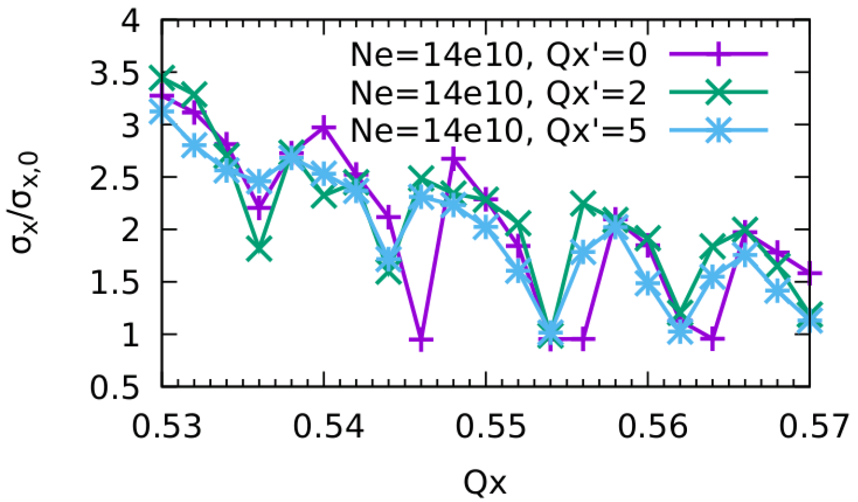}
\caption{Normalised horizontal beam size $\sigma_x/\sigma_{x,0}$ as a function of the horizontal tune for a bunch population of $N_p=1.4 \times 10^{11}$ at different chromaticities.}
\label{fig:X_Z_beta0_15_1}
\end{figure}

In Fig.~\ref{fig:X_Z_beta0_15_1} we show the results similar to those of Fig.~\ref{fig:X_Z_beta0_15}, but with a halved intensity. Also in this case it is not possible to operate the machine under stability conditions in the tune range between $Q_x=0.53 - 0.57$.

Even if we have not simulated the exact bunch population foreseen in the parameter list, we have to consider that, at this stage of the study, the bunch intensity can vary and we cannot focus on a single fixed value. The important conclusion from these simulations is that, with the present parameters, it is difficult to find a large enough stable tune area for the bunch intensities of our interest.

\section{A possible mitigation solution}
\label{s:mitigation}

In the previous section we have seen that the reduction of the single bunch intensity cannot be a solution for the problem of the X-Z instability. Also working with a positive chromaticity doesn't allow to find a stable tune area.

Another possible option that we have investigated is that of reducing the betatron function $\beta_x^*$. Indeed, as can be seen from Eq.~(\ref{eq:1}), the X-Z instability threshold is inversely proportional to the horizontal betatron function at the interaction point. This choice, however, has its downside since it also generates a reduction of the dynamic aperture and of the momentum acceptance. Therefore, a careful design of the machine optics is necessary. A good solution with a lattice having an acceptable dynamic aperture and with $\beta_x^*=10$ cm has been recently proposed~\cite{oide}.

In addition to the advantage of having a smaller betatron function at the interaction point, the parameter list with this new lattice has a lower single bunch intensity and an horizontal beam-beam parameter $\xi_x$ of about 0.0023~\cite{oide}, which is almost a factor 2 lower that that foreseen in the CDR~\cite{fccee0}. Since the width of the stable area is proportional to $\xi_x$~\cite{ohmi}, then we expect an additional stabilising effect.

\begin{figure}
\centering
\includegraphics[width=0.7\textwidth]{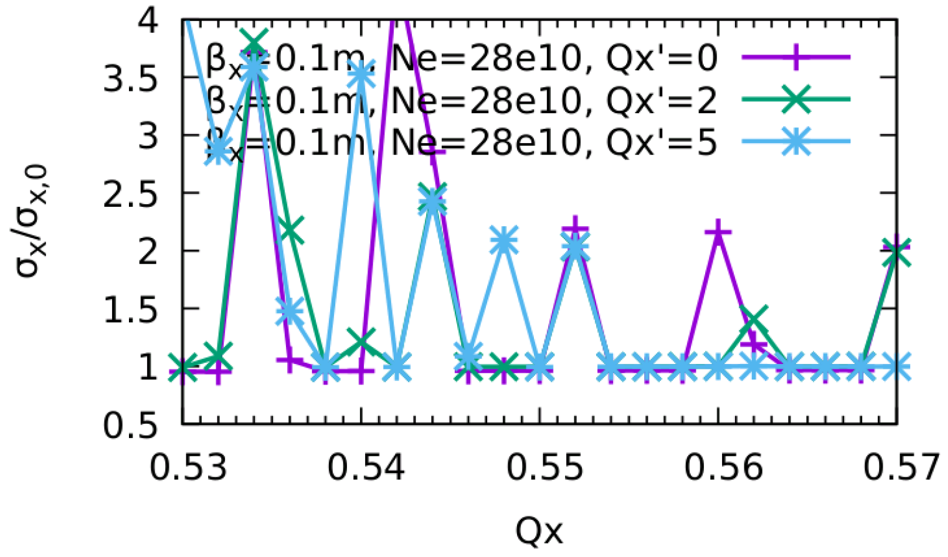}
\includegraphics[width=0.7\textwidth]{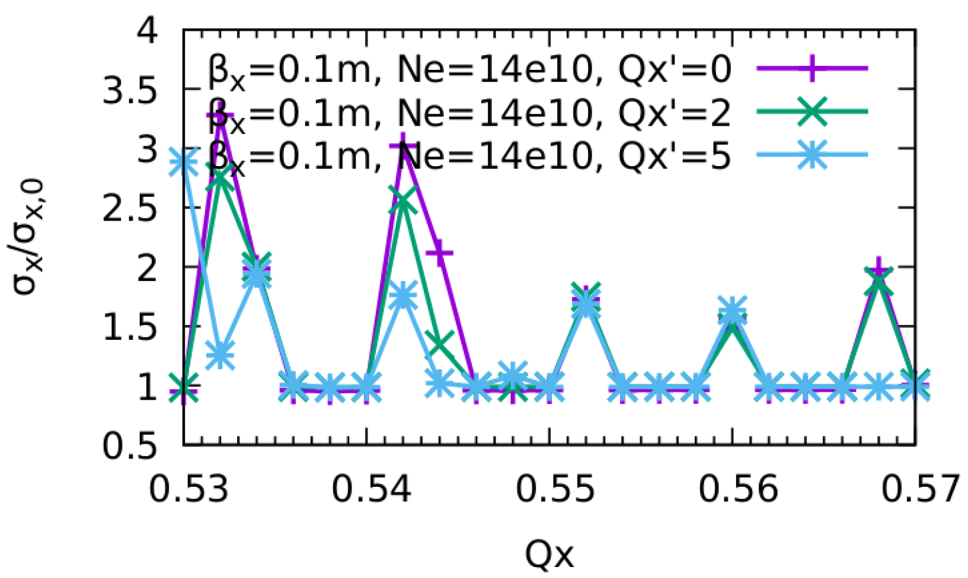}
\caption{Normalised horizontal beam size $\sigma_x/\sigma_{x,0}$ as a function of the horizontal tune for a bunch population of $N_p=2.8 \times 10^{11}$ (top) and $N_p=1.4 \times 10^{11}$ (bottom) at different chromaticities with $\beta_x^* = 10$ cm.}
\label{fig:X_Z_beta0_10}
\end{figure}

We have performed beam-beam simulations similar to those of the previous section, that is with a bunch population of $N_p = 2.8\times 10^{11}$ and $1.4\times 10^{11}$, and with chromaticities of $Q_x'=0, 2, 5$. The results are reported in Figs.~\ref{fig:X_Z_beta0_10}.

As can be seen from the two figures, the situation in this case is better than the one having $\beta_x^* = 15$ cm. Around a horizontal tune of about $Q_x =0.56$ we have two regions of stability with both intensities. These regions are even improved, at least at high intensity, with a positive value of $Q'_x$. 

\begin{figure}
\centering
\includegraphics[width=0.8\textwidth]{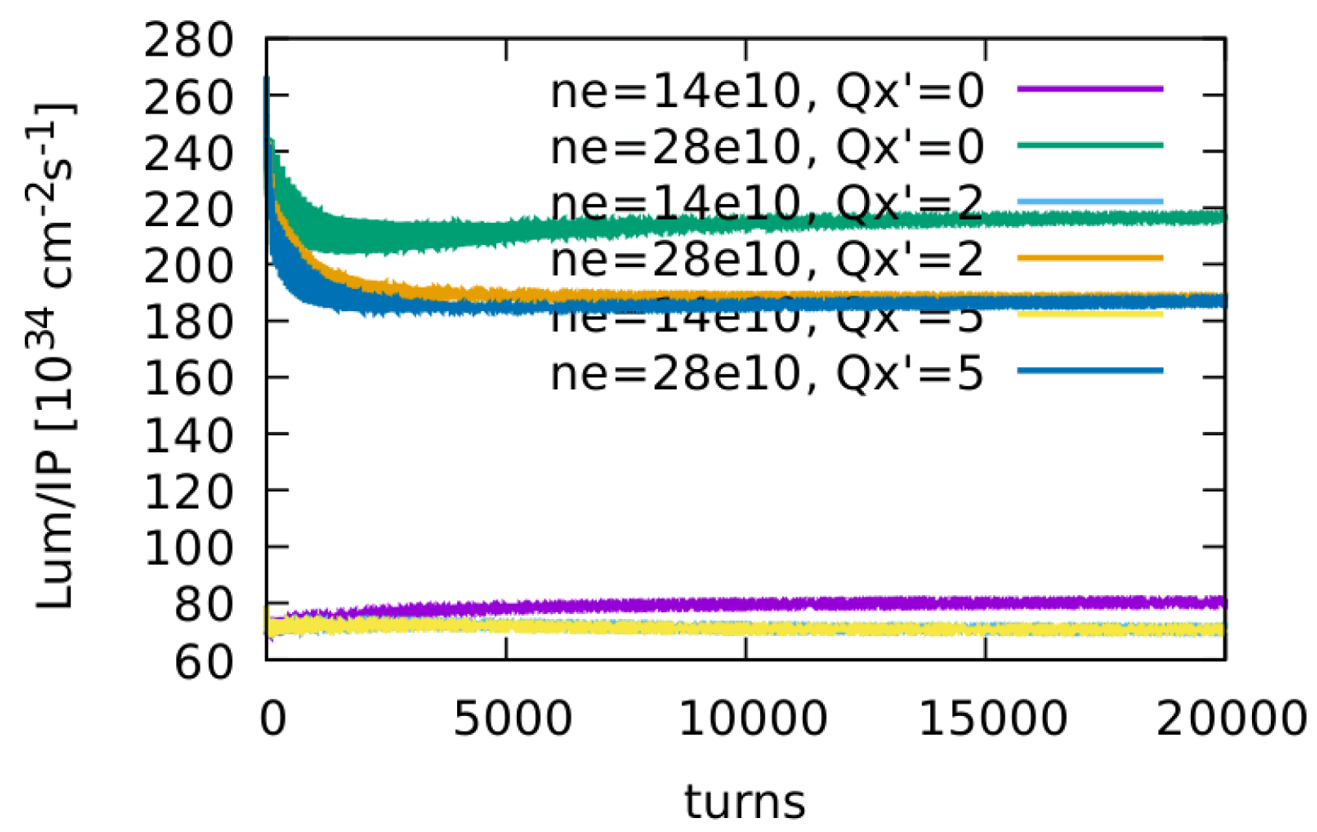}
\caption{Luminosity per IP at different chromaticities with $N_p=2.8 \times 10{11}$ and $N_p=1.4 \times 10{11}$ and with $beta_x^* = 10$ cm.}
\label{fig:luminosity}
\end{figure}

To conclude the study, we have addressed the question of how much luminosity we would loose operating the machine with this reduced $\beta_x^*$ if we take into account the longitudinal wakefield and, in case, with a chromaticity having the same values as those that we have just illustrated. In Fig.~\ref{fig:luminosity} we report the results of the simulations. 

We can see that, at the high intensity of $N_p=2.8 \times 10^{11}$, with $Q_x'=5$, we loose about 15\% of luminosity per IP, but this parameter remains above $180 \times 10^{34}$ /(cm$^2$s), as requested by the parameter list of Table~\ref{tab:1}.

\section{Conclusions}

In our previous studies it has been shown that the interplay between the longitudinal coupling impedance and the beam-beam collisions with a large Piwinski angle in FCC-ee results in a reduction of the stable betatron tune areas with respect to what was foreseen in the conceptual design report. Possible mitigation solutions investigated recently have led to an updated parameter list with a higher lattice momentum compaction factor. Furthermore, on the request of detector experts and particle physicists, a possibility of using 4 IPs instead of 2 is presently under consideration. In parallel, the machine coupling impedance model has been reviewed with more realistic representations of some devices and the evaluation of the transverse components of the wakefields. 

In this paper we have presented collective effects studies with the new parameter list relying on the updated impedance model and considering beam collisions in 4 interaction regions (4 IPs).

Without beam-beam interaction, the studies of collective effects with the contemporary presence of longitudinal and transverse wakefields show that the transverse mode coupling instability thresholds is lower than that generally evaluated by considering only the transverse plane. This is due to a synchrotron tune reduction produced by the longitudinal wakefield. The consequence is that, for example during the commissioning phase of the machine, it would not be possible to reach the nominal single bunch intensity. As for the longitudinal microwave instability, the nominal bunch intensity is only slightly above the instability threshold, and it is expected to be suppressed in collision due to the beamstrahlung that is very strong in FCC-ee.

In turn, the important coherent X-Z instability due to the beam-beam interaction has been investigated under the influence of the longitudinal wakefield for this new configuration with 4 IPs. The results show that there are no sufficiently large horizontal tune areas to operate the machine without incurring into the instability. As an eventual mitigation solution, we propose to reduce the betatron function at the IP. In particular, it has been found that, with a $\beta_x^* = 10$ cm instead of 15 cm, there are enough wide working tune areas even in collisions with 4 IPs. A positive chromaticity helps to further mitigate this instability.

The update of the impedance model will continue with the study of the collimation system, which is expected to be another important impedance source, in particular in the transverse plane. It is also in our plans to study self-consistently the 3D beam-beam interaction with 4 IPs including both the transverse and longitudinal impedances.

\section*{Acknowledgements}
The authors would like to thank the CERN vacuum group, in particular R. Kersevan and S. J. Rorison who provided the realistic model of the bellows and vacuum chamber.

%
%

\end{document}